\documentclass[nohyper,12pt,a4paper]{JHEP3}
\usepackage{epsfig,amsmath}
\usepackage{psfrag}
\usepackage{graphicx}
\usepackage{cite}

\newcommand{\half}{\frac{1}{2}}
\newcommand{\tr}{\textrm{tr}\,}

\title{Correlation functions in a $c=1$ boundary conformal field theory}

\author{Kristj\'an R.\ Kristj\'ansson$^{\ a,b}$ 
and L\'arus Thorlacius$^{\ a,c}$\\ 
a) University of Iceland, Science Institute, 
Dunhaga 3, 107 Reykjavik, Iceland\\
b) NORDITA, Blegdamsvej 17, DK-2100, Copenhagen \O, Denmark\\
c) Department of Physics, Stanford University, Stanford, CA 94305-4060, USA\\
\ \\
E-mail: \email{kristk@hi.is}, \email{lth@hi.is}} 

\abstract{
We obtain exact results for correlation functions of primary operators
in the two-dimensional conformal field theory of a scalar field 
interacting with a critical periodic boundary potential. Amplitudes 
involving arbitrary bulk discrete primary fields are given in terms
of $SU(2)$ rotation coefficients while boundary amplitudes involving 
discrete boundary fields are independent of the boundary interaction.
Mixed amplitudes involving both bulk and boundary discrete fields 
can also be obtained explicitly. Two- and three-point boundary
amplitudes involving fields at generic momentum are determined, up
to multiplicative constants, by the band spectrum in the open-string
sector of the theory.
}

\preprint{hep-th/0412175}

\begin{document}

\section{Introduction}
\label{sec:introduction}

We consider the problem of computing correlation functions of 
primary operators in a relatively simple, yet non-trivial, 
two-dimensional boundary conformal field theory. The system we study 
involves a scalar field that is free in the two-dimensional bulk but
has a periodic boundary interaction, with the period of the
interaction potential chosen such that the boundary theory preserves
half of the conformal symmetry of the bulk theory.

In the Caldeira-Leggett \cite{Caldeira:1983uj} approach to dissipative
quantum mechanics this model describes critical behaviour of a particle
subject to dissipation while moving in a periodic
potential \cite{Fisher:1985,Guinea:1985}. The model also arises in
connection with quantum Hall edge states \cite{Kane:1992} and in the
study of bosonic string theory in an on-shell open string tachyon
background \cite{Callan:1990sf}. More recently, a closely related
$c=1$ boundary conformal field theory involving a scalar field with a 
`wrong-sign' kinetic term in the bulk and a critical boundary 
potential has been applied to describe the rolling tachyon field of an 
S-brane \cite{Sen:2002nu,Strominger:2002pc,Gutperle:2003xf,Sen:2004nf}.

For all these applications it is potentially useful to supplement 
existing results on exact boundary states and partition functions by
computations of correlation functions of the primary operators in 
the theory. This task is aided by the $SU(2)$ current algebra that
resides in the model \cite{Callan:1994mw,Callan:1994ub,Polchinski:1994my}. 
For a compact boson, taking values on a circle at the self-dual radius, 
the $SU(2)$ symmetry allows us to obtain explicitly correlation
functions, that involve any bulk primary fields and/or boundary primary
fields that do not change the boundary conditions.  At other allowed 
radii we have exact results for only a subset of non-vanishing amplitudes. 

We obtain exact scaling dimensions of all primary operators in the 
boundary theory and this, along with the conformal symmetry, is sufficient
to determine low-order correlation functions up to constant coefficients.
In particular, we extend the known band structure of the system to include 
open string states where the string ends couple to different potential 
strength. The energy eigenvalues of such states enter into correlation 
functions of boundary condition changing operators. A subset of the results 
described in this paper were briefly presented in \cite{Kristjansson:2004mf}. 

The paper is organized as follows.  In the remainder of this section 
we define the model and establish our notation. We briefly
discuss some related models that have been studied in the recent 
literature. In section~\ref{bulkamplitudes} we compute scattering
amplitudes involving bulk operators. We identify the bulk discrete 
primary fields, which play a key role in what follows, and then extend
the prescription for computing bulk amplitudes of elementary string
excitations, given in \cite{Callan:1994ub}, to include bulk
discrete fields. We explicitly work out one- and two-point functions
of bulk discrete fields and verify that our prescription preserves
crossing symmetry in the interacting theory.
In section~\ref{boundaryfields} we turn our attention to boundary 
fields, including discrete boundary primary fields. We argue that 
correlation functions that involve any number of discrete boundary fields, 
and no other fields, are actually independent of the boundary coupling.
Section~\ref{openspectrum} briefly reviews the free fermion calculation 
of the open string spectrum \cite{Polchinski:1994my} and then extends
it to include open strings whose ends interact with boundary potentials
of different strength. The corresponding calculation in the closed 
string sector is carried out in an appendix. 
In section~\ref{boundarycorrelators} we discuss 
the problem of more general boundary amplitudes in models where the 
free boson is non-compact, or compactified at some multiple of the 
self-dual radius. These involve fields carrying momenta that differ from 
those of the discrete fields, and possibly also boundary condition
changing fields. The prescription we use for discrete fields 
does not carry over to the general case but conformal symmetry and the 
open string spectrum from section~\ref{openspectrum} determine low-order 
boundary correlators up to multiplicative factors, which may depend on the 
strength of the boundary coupling.  We also give a prescription for 
computing mixed amplitudes involving both bulk and boundary discrete fields.
Section~\ref{conclusions} contains a discussion and some directions
for future work.

\subsection{The system}
\label{thesystem}

Our scalar field $\Phi(z,\bar z)$ is defined on the upper-half-plane, 
$\mathrm{Im}\, z>0$, with dynamics governed by the 
action\footnote{We have set $\alpha' =2$.}
\begin{equation}
  \label{action}
  S = \frac{1}{4\pi} \int d^2 z\: \partial \Phi \bar \partial \Phi
    - \frac{1}{2} \int d\tau \left[ g e^{i\Phi(\tau)/\sqrt{2}} 
                  + \bar g e^{-i\Phi(\tau)/\sqrt{2}}\right] \,.
\end{equation}
The strength and the phase of the periodic boundary interaction is
determined by the complex parameter $g$. 
For a field that takes value on a finite circle of radius $R$ the boundary 
action in (\ref{action}) is single valued only if $2\pi R$ is an integer 
multiple of the period of the potential. This in turn means that the radius 
$R$ has to be an integer multiple of the self-dual radius of a free boson,
$R_\textrm{self-dual}=\sqrt{2}$.  
The theory at the self-dual radius itself is the simplest and our methods
yield exact results for correlation functions of arbitrary primary fields
(apart from boundary condition changing fields) in that case. We also 
consider the other allowed radii, including the the $R\rightarrow\infty$ 
limit of a non-compact scalar, but our results are less complete in those 
cases.

The period of the potential in (\ref{action}) has been 
chosen such that the interaction has dimension one under boundary
scaling. At this critical period the model is an exact boundary 
conformal field theory, {\it i.e.} half the conformal symmetry of the 
bulk system is preserved. In the process of establishing the 
conformal symmetry, the model was also found to have an underlying 
$SU(2)$ symmetry, which leads to some remarkably simple exact results
\cite{Callan:1994mw,Callan:1994ub,Polchinski:1994my}.
For example, by a judicious choice of regularization, the boundary state 
of the interacting theory was shown to be given by a global $SU(2)$ rotation,
parametrized by $g$ and $\bar{g}$, of the free boundary state and 
scattering amplitudes involving elementary string excitations were
explicitly computed \cite{Callan:1994ub}.
The open string spectrum exhibits an interesting band structure, first 
found in a free fermion representation of the model \cite{Polchinski:1994my}.
The bands are analogous to energy bands in condensed matter physics 
although the conformal symmetry of the string system leads to a spectrum 
that differs qualitatively from the standard non-relativistic case.

\subsection{Related models}
\label{relatedmodels}

Recently there has been renewed interest in $c=1$ boundary conformal
field theory and a number of models, that are related the one studied 
here, have been considered in the literature.

By formally analytically continuing the field, $\Phi\rightarrow -i\Phi$, 
and taking $g\in \mathbb{R}$ in (\ref{action}) one obtains the following 
action,
\begin{equation}  
\label{sbrane}  
S = -\frac{1}{4\pi} \int d^2 z\: \partial \Phi \bar \partial \Phi 
   - \frac{g}{2} \int d\tau \, \cosh{\Phi(\tau)/\sqrt{2}} \,,
\end{equation}
which has been applied to describe the rolling tachyon field of an S-brane
\cite{Sen:2002nu,Strominger:2002pc,Gutperle:2003xf,Sen:2004nf}. The S-brane 
instability is signalled by the `wrong sign' kinetic term in (\ref{sbrane}).
Bulk and boundary correlation functions in this unstable theory, that
describe closed and open string production in an S-brane background 
\cite{Strominger:2002pc,Gutperle:2003xf} 
are closely related to correlation functions in the static background
defined by (\ref{action}).
 
Another model studied in recent work 
\cite{Fredenhagen:2004cj,Gaberdiel:2004na} has a complex 
valued boundary potential involving only a single exponential,
\begin{equation}
  \label{complexaction}
  S = \frac{1}{4\pi} \int d^2 z\: \partial \Phi \bar \partial \Phi
    - \frac{\lambda}{2} \int d\tau \, e^{-i\Phi(\tau)/\sqrt{2}} \,.
\end{equation}
Here the boundary interaction is non-hermitian and the connection
to (\ref{action}) is more tenuous than in the case of (\ref{sbrane}).
Despite being non-hermitian, this model is in many respects simpler than 
the one defined by (\ref{action}). In either case we can view a perturbative
expansion in the boundary coupling, $g$ or $\lambda$ respectively, in terms
of a one-dimensional Coulomb gas. 
The real valued potential in (\ref{action}) corresponds to charges of both 
signs, while only same-sign charges occur in the theory with the single
exponential, and this severely restricts how the interactions contribute 
to physical quantities including correlation functions. This is for example 
reflected in the spectrum of scaling dimensions of boundary operators, which 
is independent of the coupling $\lambda$ of the single exponential 
\cite{Gaberdiel:2004na} (see also section~\ref{halfbranespectrum} below), 
but develops non-trivial dependence on $g$ when we work with the real 
valued potential in (\ref{action}).

The primary motivation for the study of the action (\ref{complexaction}) 
is its relation, by analytic continuation in the field, to a theory that 
describes the decay of an unstable D-brane in string theory,
\begin{equation}
  \label{tbl}
  S = -\frac{1}{4\pi} \int d^2 z\: \partial \Phi \bar \partial \Phi
    - \frac{\lambda}{2} \int d\tau \, e^{\Phi(\tau)/\sqrt{2}} \,.
\end{equation}
Here the $\Phi$ field is interpreted as the embedding time in a worldsheet
description of the D-brane and this model is also referred to as timelike 
boundary Liouville theory 
\cite{Gutperle:2003xf,Larsen:2002wc,Balasubramanian:2004fz}. 
In parallel with the theories with
periodic boundary interactions, the study of (\ref{tbl}) is more manageable 
than that of (\ref{sbrane}) because here the non-linear boundary term goes 
to zero at early embedding time, $\Phi\rightarrow -\infty$. 

\subsection{Left-moving $SU(2)$ current algebra}
\label{sectiontwoone}

Let us return to the study of (\ref{action}).
In the absence of the boundary interaction, the field $\Phi(z,\bar z)$
satisfies a Neumann boundary condition at $\mathrm{Im}\, z=0$, and this
is conveniently dealt with using the so-called doubling trick or method 
of images (see f.ex.\ \cite{DiFrancesco:1997nk,Polchinski:1998rq}). 
The theory on the upper half-plane is then mapped into a chiral
theory on the full complex plane and the boundary eliminated.
To see how this works we note that away from the boundary the free 
field can be written as a sum of left- and right-moving (holomorphic and 
anti-holomorphic) components
\begin{equation}
\Phi(z,\bar z) = \phi(z) + \bar \phi(\bar z).
\end{equation}
For $g=0$ the Neumann boundary condition on the real axis 
$\bar \phi(\bar z) - \phi(z)\vert_{z=\bar z} = 0$ 
determines the right-moving field in terms
of the left-moving one. We can therefore reflect the right-moving field 
through the boundary and represent it as a left-moving field in the 
lower-half-plane.  The theory then contains only left-moving fields but
a left-moving field at $z^*$ in the unphysical lower-half-plane is to be 
interpreted as a right-moving field at $z$ in the physical 
region\footnote{Our notation follows that of \cite{DiFrancesco:1997nk}.
Both $\bar z$ and $z^*$ denote the complex conjugate of $z$.  We use
$\bar z$ for the argument of a right-moving, anti-holomorphic field 
$\bar\phi(\bar z)$, but $z^*$ for that of the corresponding left-moving 
image field $\phi(z^*)$.}.

More generally, any quasi-primary field in the bulk separates into a 
holomorphic and an anti-holomorphic part
\begin{equation}
\Psi_{h,\bar h}(z,\bar z) = \psi_h(z)\bar \psi_{\bar h}(\bar z).
\end{equation}
Using the doubling trick 
$\bar \psi_{\bar h}(\bar z)$ becomes a holomorphic field 
$\psi_{\bar h}(z^*)$, with holomorphic dimension $\bar h$.
A bulk $n$-point function 
$\langle \psi_{h_1,\bar h_1}(z_1,\bar z_1) \cdots
\psi_{h_n,\bar h_n}(z_n,\bar z_n)\rangle$ 
in the original theory on the upper-half-plane then becomes a
$2n$-point function of holomorphic fields
$\langle \psi_{h_1}(z_1)\psi_{\bar h_1}(z_1^*) \cdots
\psi_{h_n}(z_n)\psi_{\bar h_n}(z_n^*)\rangle$ 
on the infinite plane.

It turns out that the doubling trick can be applied even when
the boundary interaction in (\ref{action}) is turned on.  This is 
because the boundary potential can in fact be expressed in terms of 
left-moving fields only as
\begin{equation}
\label{bpotential}
-\frac{1}{2} 
\left( g\, e^{i\sqrt{2}\phi(z)} + \bar g\, e^{-i\sqrt{2}\phi(z)}\right)
\bigg\vert_{\mathrm{Im}(z) = 0}
\end{equation}
Note that the coefficient in the exponential has changed because 
on the real axis $\Phi(z,\bar z) = 2\phi(z)$.  

The operators appearing in the interaction (\ref{bpotential}) are 
currents of a left-moving $SU(2)$ algebra
\begin{equation}
\label{currents}
J_\pm = e^{\pm i\sqrt{2}\phi(z)}, \qquad 
J_3 = \frac{i}{\sqrt{2}}\partial \phi(z).
\end{equation}
In the boundary action (\ref{action}) the $J_+$ and $J_-$ currents are 
integrated along the real axis (the former boundary) and in a perturbative 
expansion of a bulk correlation function such integrals are repeatedly 
inserted into the amplitude.   As usual, divergences arise when operator 
insertions coincide but, by a clever choice of regularization, Callan 
{\it et al.}\ \cite{Callan:1994ub} were able to sum the perturbation
series explicitly to obtain the exact interacting boundary state.  It
turned out to be remarkably simple, with the net effect of the interaction 
being a global $SU(2)$ rotation, 
\begin{equation}
\label{groupelement}
U(g_r) = \exp \pi i (g_r J_+ + \bar g_r J_-) \,,
\end{equation}
acting on the free Neumann boundary state. The coupling constants $g_r$ 
and $\bar g_r$ that enter in the rotation group element are renormalized
from their bare values \cite{Callan:1994ub}. 
In the weak coupling limit $g_r$ approaches $g$, whereas at
strong coupling when $\vert g\vert \rightarrow \infty$ the renormalized
coupling instead goes to a finite value, $\vert g_r\vert\rightarrow 1/2$. 
The explicit form of
this redefinition has recently been worked out to be \cite{Kogetsu:2004te}
\begin{align}
\label{renormg}
g_r = \frac{2g}{\pi\vert g\vert} 
 \arctan\left[\tanh\left(\frac{\pi}{2}\vert g\vert\right)\right].
\end{align}
It is the renormalized coupling that enters in our formulas below and we
will drop the $r$ subscript on $g_r$ henceforth.

\section{Bulk scattering amplitudes}
\label{bulkamplitudes}

In this section we consider scattering amplitudes involving bulk fields.
General bulk amplitudes are functions of the boundary coupling $g$ and 
our goal is to determine this dependence.  Some bulk amplitudes that are 
zero in the free theory are nonvanishing in the presence of the boundary 
interaction.  This is because the boundary interaction breaks translation 
invariance in the target space.  It represents a periodic background that 
can absorb momenta lying in the corresponding reciprocal lattice. 

\subsection{Bulk primary fields}
\label{bulkprimaries}

The bulk theory is that of a free boson.  For a non-compact boson there are
holomorphic primary fields $e^{ip\phi(z)}$ with conformal weight $h=p^2/2$ 
for all $p\in \mathbb{R}$ and also the corresponding anti-holomorphic fields.  
If the boson instead takes value on a circle of finite radius $R$, then the
momentum variable is discrete, $p = n/R$ with $n\in \mathbb{Z}$.
Recall that the only radii that are commensurate with the period of
the boundary potential are integer multiples of the self-dual radius
$R_\textrm{self-dual}=\sqrt{2}$.

At special values of the momentum, $p = \sqrt{2}\,j$, where $j$ is an 
integer or integer-plus-half, some descendant states have vanishing norm 
and new primary fields appear, the so-called discrete primaries 
\cite{Kac:1979}. 
We note that these special values of momentum coincide with the 
allowed momenta at the self-dual radius and are among the allowed 
momenta at any integer multiple of $R_\textrm{self-dual}$.

The discrete primary fields come in $SU(2)$ multiplets labelled by $j$ and 
$m$, with $-j \le m \le j$.  The discrete fields $\psi_{jm}(z)$ in a given 
$SU(2)$ multiplet are degenerate in that they all have the same holomorphic 
conformal weight $h=j^2$.  They are composite fields made from certain 
polynomials in $\partial\phi$, $\partial^2\phi$, {\it etc.} accompanied by 
$e^{i\sqrt{2}m\phi}$. 

The $\psi_{jm}$ have the following representation \cite{Klebanov:1991hx},
\begin{equation}
\label{discretefields}
\psi_{jm}(z) \sim
\left(\oint \frac{dw}{2\pi i} e^{-i\sqrt{2}\phi(w)}\right)^{j-m}
e^{i\sqrt{2}j\phi(z)} \,,
\end{equation}
where the lowering current is integrated along nested contours 
surrounding $z$.  The operator products are evaluated using Wick 
contractions and the free holomorphic propagator 
\begin{equation}
\label{freeprop}
\langle \phi(z)\phi(z')\rangle = - \log (z-z') \,.
\end{equation}
This propagator is also used in normal ordering composite fields, such as
those that appear in (\ref{discretefields}), and we refer to
this as {\it bulk normal ordering}.

We adopt the convention
\begin{equation}
\psi_{j,m-1}(z) = c_{j,m} \,
\oint \frac{dw}{2\pi i} e^{-i\sqrt{2}\phi(w)}\, \psi_{jm}(z),
\end{equation}
with $c_{j,m}=\left[j(j+1)-m(m-1)\right]^{-1/2}$, for the relative 
normalization of the $\psi_{jm}$ and the overall normalization is fixed
by setting $\psi_{jj} = e^{i\sqrt{2}\,j\phi}$.\footnote{Strictly
speaking we need to include a cocycle in the definition of $\psi_{jj}$
in order to ensure locality of the OPE involving $\psi_{jm}$'s with $j$ an 
integer plus a half (see for example \cite{Polchinski:1998rq}, page 239). 
Such a cocycle will affect the overall sign of some amplitudes 
but to avoid clutter we leave it out.}

Let us list a few explicit examples for later use. The $j=0$ case
is trivial, $\psi_{0,0}=1$, and at $j=1/2$ there is a doublet,
\begin{equation}
\psi_{\half,\half} = e^{i\phi/\sqrt{2}}, \qquad 
\psi_{\half,-\half} = e^{-i\phi/\sqrt{2}}. 
\end{equation}
At $j=1$ we recognize the $SU(2)$ currents themselves, modulo constant
factors,
\begin{equation}
\psi_{1,1} = e^{i\sqrt{2}\phi}, \quad
\psi_{1,0} = -i  \partial\phi, \quad
\psi_{1,-1} = -e^{-i\sqrt{2}\phi},
\end{equation}
and at $j=3/2$ we find
\begin{align}
\psi_{\frac{3}{2},\frac{3}{2}}& = e^{3i\phi/\sqrt{2}}, &
\psi_{\frac{3}{2},\frac{1}{2}}& =\frac{-1}{\sqrt{3}}
\left((\partial\phi)^2 + \frac{i}{\sqrt{2}}\partial^2\phi\right)
  e^{i\phi/\sqrt{2}}, \nonumber \\
\psi_{\frac{3}{2},-\frac{3}{2}}& = -e^{-3i\phi/\sqrt{2}}, &
\psi_{\frac{3}{2},-\frac{1}{2}}& = \frac{1}{\sqrt{3}}
\left((\partial\phi)^2 - \frac{i}{\sqrt{2}}\partial^2\phi \right)
  e^{-i\phi/\sqrt{2}} \,. 
\end{align}

The operator product expansion (OPE) of holomorphic discrete fields 
defines a closed operator algebra,
\begin{align}
\label{eq:chiralope}
\psi_{jm}(z)\psi_{j'm'}(w) = \sum_{J,M}
\frac{A^{J M}_{jm, j'm'} \, \psi_{J M}(w) }{(z-w)^{j^2+j'^2 -J^2}}
+ \cdots \,,
\end{align}
where the $A^{JM}_{jm, jm'}$ are constant OPE coefficients and the 
ellipsis denotes the contribution from descendants of the 
$\psi_{JM}$ primaries. The $SU(2)$ symmetry is apparent in the 
OPE coefficients,
\begin{align}
\label{eq:AfC}
A^{J M}_{jm, j'm'} = f(j,j',J) \, C^{J M}_{jm, j'm'}
\end{align}
where $C^{J M}_{jm, jm'}$ are Clebsch-Gordan coefficients and 
$f(j,j',J)$ are numbers that can be obtained by direct 
calculation \cite{Klebanov:1991hx}. The Clebsch-Gordan coefficients 
restrict the sum over $J$ in (\ref{eq:chiralope}) to 
$\vert j-j'\vert \le J \le j+j'$ and the only value of $M$ that
contributes is $M=m+m'$.
For $j=1$ and $j'=1/2$ we have for example
\begin{align}
\psi_{11}(z)\psi_{\half \half}(w) 
 &= (z-w)\ \psi_{\frac{3}{2}\frac{3}{2}}(w) + {\cdots} \\
\psi_{10}(z)\psi_{\half \half}(w) 
 &= \frac{-1/\sqrt{2}}{(z-w)}\, \psi_{\half\half}(w) 
 - \sqrt{2}\, \partial\psi_{\half\half}(w) \nonumber\\
 &+ (z-w)\left[ \sqrt{\frac{2}{3}}\, \psi_{\frac{3}{2}\half}(w) 
       -\frac{2\sqrt{2}}{3}\, \partial^2 \psi_{\half\half}(w)\right]
 + {\cdots}
\end{align}
From the first OPE we see that $f(1,\half,\frac{3}{2})=1$ and in
the second one the coefficient in front of
$\psi_{\frac{3}{2}\half}$ is $\sqrt{2/3}$ which is the
expected Clebsch-Gordan coefficient. 

The bulk theory also contains anti-holomorphic discrete fields, which also
form $SU(2)$ multiplets and have their own OPE. A {\it bulk discrete primary 
field\/} is constructed from a pair of holomorphic and anti-holomorphic 
discrete fields,
\begin{equation}
\label{bulkdiscreteprimary}
\Psi_{j m, \bar\jmath \bar m}(z,\bar z) =
\psi_{j m}(z)\bar\psi_{\bar\jmath \bar m}(\bar z) \,.  
\end{equation}
The left- and right-moving parts can carry different $SU(2)$ 
labels subject to the following restrictions. 

Let us first consider a compact boson at the self-dual radius.
The winding number around the compact circle is given by 
$w=m-\bar m$ and this requires 
\begin{equation}
m-\bar m\in \mathbb{Z} \,. 
\end{equation}
It follows that the difference $j-\bar\jmath$ must also be an integer.
We note also that the spin $h-\bar h=j^2-\bar\jmath^2$ takes 
an unphysical value unless $j-\bar\jmath\in \mathbb{Z}$. 

The target space momentum carried by the field (\ref{bulkdiscreteprimary})
is $p_{m,\bar m} = (m+\bar m)/\sqrt{2}$.  The allowed momentum values at 
the self-dual radius, $R_\textrm{self-dual}=\sqrt{2}$, correspond precisely 
to the integer values taken by $m+\bar m$. This means that, at the self-dual 
radius, the bulk primary fields in the theory consist of the bulk 
discrete primaries, and no others. As a result all bulk correlation 
functions in the theory at the self-dual radius can be computed by the 
$SU(2)$ methods discussed below.

If the boson is compactified at some integer multiple of the 
self-dual radius, $R=q\,\sqrt{2}$ with $q\in \mathbb{Z}$, the restriction 
from the winding number becomes more stringent
\begin{equation}
m-\bar m = 0 \mod q \,,
\end{equation}
but at the same time there are additional bulk primary fields in the 
theory that carry momenta  $p=n/(q\sqrt{2})$ with $n\in \mathbb{Z}$.
The operator product of the $SU(2)$ currents (\ref{currents})
with these additional primary fields is non-local. 

In the limit of a non-compact boson there can be no winding number, which 
amounts to the restriction $\bar m=m$ on the bulk discrete primary fields.
On the other hand, the target space momentum of a non-compact boson is 
unrestricted leading to a continuum of bulk primary fields
\begin{equation}
\label{bulkprimary}
\Psi_p(z,\bar z) =
e^{ip(\phi(z)+\bar\phi(\bar z))} \,,
\end{equation}
with $p\in \mathbb{R}$.

\subsection{Method of rotated images}

Callan {\it et al.} \cite{Callan:1994ub} gave a simple prescription for 
amplitudes that describe the scattering of elementary string excitations 
off the interacting worldsheet boundary. Let us briefly review this 
prescription, which we refer to as {\em the method of rotated images}.

Left-moving string excitations are created and destroyed by 
$\partial \phi$ so for each incoming field a factor of $\partial \phi(z)$ 
is inserted in the upper-half-plane.  For each right-moving outgoing 
excitation we normally would insert $\bar\partial\bar\phi(\bar z)$ also 
above the real axis, but using the doubling trick we instead insert the 
reflected operator $\partial\phi(z^*)$ in the lower-half-plane.  
This is possible even with the boundary interaction
turned on because the anti-holomorphic fields $\bar\partial\bar\phi$ 
commute with the holomorphic $SU(2)$ currents in the interaction and 
reflect through the underlying Neumann boundary condition exactly as in
the free theory.

The integration contours of the $SU(2)$ currents that appear in the boundary 
interaction can now be deformed into the lower half-plane and moved off to
infinity. This leaves behind closed integration contours around each image
field in the lower half-plane and, since the $\partial\phi(z^*)$ fields 
are themselves proportional to left-moving $SU(2)$ currents, 
the current algebra ensures that there will be no cuts generated.
The terms in the perturbative series involve an ever higher number
of nested contours surrounding each $\partial\phi(z^*)$ field insertion, but 
these sum up to very simple result: a global $SU(2)$ rotation acting on 
each inserted field.  The global $SU(2)$ element is given by 
(\ref{groupelement}), which for $g = \bar g$ amounts to a rotation by the 
angle $2\pi g$ about the $J_1$ axis, giving 
\begin{equation}
\label{dphirotated}
\partial\phi(z^*) \rightarrow \cos(2\pi g)\,\partial\phi(z^*)
-\frac{\sin(2\pi g)}{\sqrt{2}} 
\left(e^{i\sqrt{2}\phi(z^*)}-e^{-i\sqrt{2}\phi(z^*)}\right) \,.
\end{equation}
Finally one evaluates the resulting correlator of left-moving fields using
the free holomorphic propagator (\ref{freeprop}). 

\subsection{Bulk amplitudes involving discrete primary fields}

The method of rotated images can be generalized to compute scattering 
amplitudes involving arbitrary bulk discrete fields. The key to this 
is the fact that the boundary interaction (\ref{bpotential}) involves 
left-moving $SU(2)$ currents which act on the discrete fields in a 
well defined manner.  As discussed above, for a compact boson at the 
self-dual radius, all bulk primary fields are discrete fields so the 
method can be applied to all bulk amplitudes of interest in that model. 
At other allowed radii there are 
non-vanishing amplitudes to which the method cannot be applied.
In this case, bulk amplitudes can involve both the discrete bulk fields 
and fields carrying generic target space momenta.
The only constraint from momentum conservation is that the total momentum
of all the fields in a given correlator add up to an integer multiple of
the lattice momentum of the periodic background boundary potential.
The operator product of the currents in the boundary interaction and 
generic momentum fields is non-local so the effect of the interaction 
is no longer captured by a global $SU(2)$ rotation.  The following 
calculations therefore only apply to the subset of bulk amplitudes 
where all the operator insertions involve discrete fields. 

A general bulk discrete primary field is given by 
(\ref{bulkdiscreteprimary}).
Insert one of these fields in the upper half-plane and use
the method of images so that
$\bar\psi_{\bar\jmath \bar m}(\bar z) \to \psi_{\bar\jmath \bar m}(z^*)$.
As before, the boundary interaction acts on the image operator in the 
lower-half-plane by the global $SU(2)$ rotation $U(g)$ given in 
(\ref{groupelement}).  The original right-moving discrete field was a 
component of a rank $\bar\jmath$ irreducible tensor operator, so the 
rotated image is
\begin{equation}
\widetilde \psi_{\bar\jmath \bar m}(z^*)
= \sum_{\bar m'=-\bar\jmath}^{\bar\jmath} 
\mathcal{D}^{\bar\jmath}_{\bar m,\bar m'}(g) 
\psi_{\bar\jmath \bar m'}(z^*) \,,
\end{equation}
with the rotation coefficient given by
\begin{equation}
\mathcal{D}^{\bar\jmath}_{\bar m,\bar m'}(g)=
\langle \bar\jmath,\bar m\vert U(g) \vert \bar\jmath,\bar m'\rangle
= \langle \bar\jmath,\bar m\vert 
e^{\pi i (g J_+ + \bar g J_-)} \vert \bar\jmath,\bar m'\rangle \,,
\label{eq:coeff}
\end{equation}
where $\vert j,m\rangle$ are standard $SU(2)$ states.
For simplicity we will choose $g$ to be a real number in the following.
In that case the coefficients $\mathcal{D}^{j}_{m,m'}(g)$
can be expressed in terms of $\cos\pi g$ and $\sin\pi g$.
A general formula is given in appendix \ref{app:rotation}, along with 
some explicit examples.

A bulk $n$-point function involving bulk discrete primary fields in the
interacting theory can  therefore be expressed in terms of $2n$-point 
functions of holomorphic discrete fields in the free theory along 
with  $SU(2)$ rotation coefficients,
\begin{eqnarray}
\label{bulknpoint}
\langle \Psi_{j_1 m_1, \bar\jmath_1 \bar m_1}(z_1,\bar z_1)&\ldots&
\Psi_{j_n m_n, \bar\jmath_n \bar m_n}(z_n,\bar z_n)\rangle_g \\
&=& \sum_{\bar m'_i=-\bar\jmath_i}^{\bar\jmath_i} 
\mathcal{D}^{\bar\jmath_1}_{\bar m_1,\bar m'_1}(g)
\ldots \mathcal{D}^{\bar\jmath_n}_{\bar m_n,\bar m'_n}(g)\,
\langle \psi_{j_1 m_1}(z_1)\psi_{\bar\jmath_1 \bar m'_1}(z_1^*)
\ldots\rangle \,.
\nonumber
\end{eqnarray}
The only dependence on the boundary coupling is through the rotation 
coefficients that appear in (\ref{bulknpoint}).
\psfrag{psi1}{$\psi_{jm}(z)$}
\psfrag{tpsi1}{$\widetilde\psi_{\bar\jmath\bar m}(z^*)$}
\psfrag{Psi1}{$\Psi_{jm,\bar\jmath\bar m}(z,\bar z)$}
\EPSFIGURE{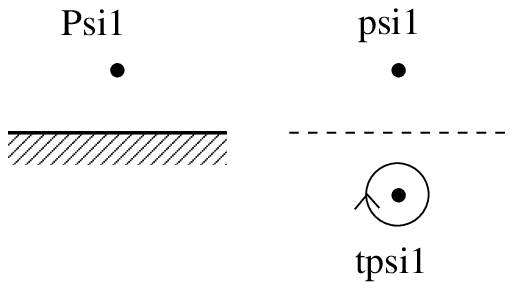, width=6cm}{
The method of rotated images applied to the one-point 
function of a bulk discrete primary field.
\label{fig:methrotim}}

Below we will illustrate this general result by considering one- and 
two-point functions of bulk fields, but let us first note that these 
bulk correlators have another equivalent representation.  
When implementing the method of rotated images we chose to deform the 
integration contours of the boundary currents into the lower half-plane 
but we could just as well have moved them into the upper half-plane 
instead. This results in the inverse $SU(2)$ rotation acting on the 
left-moving part of each bulk field,
\begin{equation}
\widetilde \psi_{jm}(z)
= \sum_{m'=-j}^j
\mathcal{D}^j_{m,m'}(-g) \psi_{jm'}(z) \,,
\end{equation}
and the following expression for the bulk $n$-point function,
\begin{eqnarray}
\langle \Psi_{j_1 m_1, \bar\jmath_1 \bar m_1}(z_1,\bar z_1)&\ldots&
\Psi_{j_n m_n, \bar\jmath_n \bar m_n}(z_n,\bar z_n)\rangle_g \\
&=& \sum_{m'_i=-j_i}^{j_i}
\mathcal{D}^{j_1}_{m_1,m'_1}(-g)
\ldots \mathcal{D}^{j_n}_{m_n,m'_n}(-g)\,
\langle \psi_{j_1 m'_1}(z_1)\psi_{\bar\jmath_1 \bar m_1}(z_1^*)
\ldots\rangle \,.
\nonumber
\end{eqnarray}
At first glance this expression looks different from (\ref{bulknpoint})
but the two must be equivalent. We will verify this for bulk one-point
functions below and we have also checked that it works for a 
number of examples involving higher-point functions.

\subsubsection{Bulk one-point functions}
\label{bulkonepoint}

In the free theory with Neumann boundary conditions momentum
conservation implies that the only bulk operator that has a 
non-vanishing one-point function is the unit operator.  The
periodic boundary interaction can absorb momenta $p=n/\sqrt{2}$,
with $n\in \mathbb{Z}$, 
and so any bulk operator that carries such a momentum will 
have a non-vanishing one-point function.  These are precisely
the discrete bulk operators and the method of rotated images 
gives
\begin{equation}
\langle \Psi_{j m, \bar\jmath \bar m}(z,\bar z)\rangle_g =
\sum_{\bar m'=-\bar\jmath}^{\bar\jmath} 
\mathcal{D}^{\bar\jmath}_{\bar m,\bar m'}(g)
\langle \psi_{j m}(z) \psi_{\bar\jmath \bar m'}(z^*) \rangle \,.
\end{equation}
Conformal invariance requires the scaling dimension of the two chiral 
operators to be the same, i.e. $\bar\jmath^2 = j^2$, while momentum 
conservation in the free chiral theory requires $\bar m'=-m$.  
The bulk one-point function is therefore given by 
\begin{eqnarray}
\langle \Psi_{j m, \bar\jmath \bar m}(z,\bar z) \rangle_g
&=&\delta_{j,\bar\jmath}\,\mathcal{D}^{j}_{\bar m,-m}(g) \,
\langle \psi_{j,m}(z) \psi_{j,-m}(\bar z) \rangle
\nonumber \\
&=& (-1)^{N[j,m]} \, \delta_{j,\bar\jmath}\, 
\frac{\mathcal{D}^{j}_{\bar m,-m}(g)}{(z-z^*)^{2j^2}} \,,
\label{oneptfcn}
\end{eqnarray}
where $N[j,m]$ is an integer that depends on the normalization conventions 
for the discrete fields.  We can relate the $N[j,m]$ for different
values of $m$ belonging to any given $j$ as follows. The chiral two-point 
function in \eqref{oneptfcn} involves two discrete fields whose operator
product includes $\psi_{00}$, also known as the unit operator, with an 
OPE coefficient that is proportional to the Clebsch-Gordan coefficient
$C^{00}_{j,m;j,-m}$. For a given value of $j$ these particular 
Clebsch-Gordan  coefficients all have the same magnitude but alternate in 
sign with $m$,
\begin{equation}
C^{00}_{j,m;j,-m} = (-1)^{j-m} C^{00}_{j,j;j,-j} \,.
\label{cgproperty}
\end{equation}
These alternating signs are inherited by the $N[j,m]$ for a given $j$,
leaving only $N[j,j]$ to be determined case by case.

If we instead move the integration contours of the boundary currents into 
the upper half-plane we obtain an answer that looks somewhat different,
\begin{equation}
\langle \Psi_{j m, \bar\jmath \bar m}(z,\bar z)\rangle_g =
(-1)^{N[j,-\bar m]} \,
\delta_{j,\bar\jmath}\,
\frac{\mathcal{D}^{j}_{m,-\bar m}(-g)}{(z-z^*)^{2j^2}} \,,
\end{equation}
but due to the property, 
$\mathcal{D}^{j}_{m,-\bar m}(-g)=(-1)^{-(m+\bar m)}
\mathcal{D}^{j}_{\bar m,-m}(g)$, 
which follows from the general formula for the rotation 
coefficients given in appendix \ref{app:rotation}, and the fact 
that $N[j,-\bar m]=(-1)^{m+\bar m} N[j,m]$, which follows from
\eqref{cgproperty}, the two expressions for the bulk one-point function 
are in fact exactly the same.

\subsubsection{Bulk two-point functions}
A two-point function of bulk discrete fields involves rotation coefficients
and four-point functions of holomorphic discrete fields, which in turn can 
be expressed in terms of OPE coefficients and holomorphic conformal blocks.
Let's consider the two-point function of generic bulk discrete primaries,
\begin{align}
\label{bulktwopoint}
\langle
\Psi_{j_1 m_1, \bar\jmath_1 \bar m_1}(z,\bar z)
\Psi_{j_2 m_2, \bar\jmath_2 \bar m_2}(w, \bar w)
\rangle_g \,,
\end{align}
which according to the method of rotated images has
the chiral form 
\begin{align}
\label{chiralform}
\sum_{\bar m_1'=-\bar\jmath_1}^{\bar\jmath_1}
\sum_{\bar m_2'=-\bar\jmath_2}^{\bar\jmath_2}
\mathcal{D}^{\bar \jmath_1}_{\bar m_1,\bar m_1'}
\mathcal{D}^{\bar\jmath_2}_{\bar m_2,\bar m_2'}
\langle
\psi_{j_1 m_1}(z)\psi_{\bar\jmath_1 \bar m_1'}(z^*)
\psi_{j_2 m_2}(w)\psi_{\bar\jmath_2 \bar m_2'}(w^*)
\rangle \,.
\end{align}
The chiral four-point functions can in principle be evaluated by using 
the free holomorphic propagator \eqref{freeprop}.

We can express any bulk two-point function of discrete primary 
fields in terms of chiral $SU(2)$ conformal blocks.
Each chiral four-point function that appears in \eqref{chiralform} can be 
reduced to a sum over two-point functions by first applying the holomorphic 
OPE \eqref{eq:chiralope} to $\psi_{j_1 m_1}(z)$ and the image field
$\psi_{\bar\jmath_1 \bar m_1'}(z^*)$ and separately to 
$\psi_{j_2 m_2}(w)$ and $\psi_{\bar\jmath_2 \bar m_2'}(w^*)$. 
The end result can be expressed in terms of conformal blocks 
$\mathcal{F}^{(j_1 m_1)(\bar\jmath_1 \bar m_1')}_{(j_2 m_2)
(\bar\jmath_2 \bar m_2')}(J \vert \eta)$
with intermediate states belonging to the conformal
family of highest weight $J^2$, where 
$\vert j_1-\bar\jmath_1\vert\le J \le j_1+\bar\jmath_1$.
Each conformal block depends on the position of
the operators only through the anharmonic ratio 
\begin{equation}
\eta = \frac{(z-z^*)(w-w^*)}{(z-w^*)(w-z^*)} \,,
\end{equation}
which is real valued, $0\leq\eta\leq 1$, for all $z,w$ in the
upper half plane. The sum over conformal blocks is
accompanied by a prefactor involving powers of $(z-w)$, $(z-w^*)$,
{\it etc.}, that has the required scaling dimension 
$j_1^2+\bar\jmath_1^2+j_2^2+\bar\jmath_2^2$.
We use standard conventions for conformal blocks \cite{Belavin:1984vu}.  
The form of the prefactor is fixed by the mapping that takes
the standard arguments of the fields in the four-point function 
$(0,\eta,1,\infty)$ to the ones we are using $(\bar z,z,w,\bar w)$.
The bulk two-point function (\ref{bulktwopoint}) is then given by
\begin{align}
\label{eq:genbulk2pt}
(z-w^*)&^{-2j_1^2}
(w-z^*)^{-j_1^2 -\bar\jmath_1^2-j_2^2+\bar\jmath_2^2}
(w-w^*)^{j_1^2 +\bar\jmath_1^2-j_2^2-\bar\jmath_2^2}
(z^*-w^*)^{j_1^2-\bar\jmath_1^2+j_2^2-\bar\jmath_2^2}
\nonumber \\
&\times
\sum_{\bar m_1', \bar m_2'}
\mathcal{D}^{\bar\jmath_1}_{\bar m_1,\bar m_1'}
\mathcal{D}^{\bar\jmath_2}_{\bar m_2,\bar m_2'}
\sum_{J}
A^{J (m_1+\bar m_1')}_{j_1 m_1, \bar\jmath_1 \bar m_1'}
A^{J (m_2+\bar m_2')}_{j_2 m_2, \bar\jmath_2 \bar m_2'}
\mathcal{F}^{(j_1 m_1)(\bar\jmath_1 \bar m_1')}_{(j_2 m_2)
(\bar\jmath_2 \bar m_2')} (J \vert \eta) \,.
\end{align}
The conformal block 
$\mathcal{F}^{(j_1 m_1)(\bar\jmath_1 \bar m_1')}_{(j_2 m_2)
(\bar\jmath_2 \bar m_2')}(J \vert \eta)$
is zero if $m_1+\bar m_1'+m_2+\bar m_2'\neq 0$
but since we sum over $\bar m_1'$ and $\bar m_2'$ there will in general 
be a non-vanishing contribution to the sum in \eqref{eq:genbulk2pt}.

As an example of a bulk two-point function we take 
$j_i = \bar \jmath_i =\half$ and
$m_1=\bar m_1=\half, m_2=\bar m_2=-\half$.
The bulk fields decompose as
\begin{align}
\Psi_{\half\half,\half\half}(z,\bar z)&
=\psi_{\half \half}(z) \; \bar \psi_{\half \half}(\bar z)\,, \\
\Psi_{\half(-\half),\half(-\half)}(w,\bar w)&
=\psi_{\half(-\half)}(w) \; \bar \psi_{\half(-\half)}(\bar w) \,,
\end{align}
and the rotated image fields are
\begin{align}
\tilde \psi_{\half \half}(z^*) &=
\sum_{m=\pm\half} \mathcal{D}^\half_{\half, m}(g)\;\psi_{\half m} (z^*)
=\cos \pi g\; \psi_{\half \half}(z^*)+i\sin \pi
g\;\psi_{\half(-\half)}(z^*) \,, \nonumber\\
\tilde \psi_{\half (-\half)}(w^*)
&= \sum_{m=\pm\half} \mathcal{D}^\half_{-\half,m}(g)\;\psi_{\half m} (w^*)
= i\sin \pi g\; \psi_{\half \half}(w^*)+\cos \pi g\;
\psi_{\half (-\half)}(w^*) \,. \nonumber
\end{align}
There are four chiral four-point functions to be evaluated but only
two of them conserve momentum and we get
\begin{align}
\langle\Psi_{\half\half,\half\half}&(z, \bar z)
\Psi_{\half(-\half),\half(-\half)}(w,\bar w)\rangle_g \\
=& \cos^2 \pi g\; \left\langle
\psi_{\half \half}(z)\;\psi_{\half \half}(z^*)\;
\psi_{\half (-\half)}(w)\;\psi_{\half (-\half)}(w^*)
\right\rangle \nonumber \\
&-\sin^2 \pi g\; \left\langle
\psi_{\half \half}(z)\;\psi_{\half (-\half)}(z^*)\;
\psi_{\half (-\half)}(w)\;\psi_{\half \half}(w^*)
\right\rangle \nonumber \\
=&
\frac{\cos^2 \pi g}{|z-w|}
\left[\frac{(z-z^*)(w-w^*)}{(z-w^*)(z^*-w)}\right]^{1/2}
-\frac{\sin^2 \pi g}{|z-w|}
\left[\frac{(z-z^*)(w-w^*)}{(z-w^*)(z^*-w)}\right]^{-1/2} \,.
\nonumber
\end{align}
This can be reexpressed to match the general formula 
\eqref{eq:genbulk2pt} as follows, 
\begin{align}
\langle\Psi&_{\half\half,\half\half}(z, \bar z)
\Psi_{\half(-\half),\half(-\half)}(w,\bar w)\rangle_g \\
=& \frac{1}{(z-w^*)^{1/2}(w-z^*)^{1/2}} 
\left(\cos^2 \pi g\;\left[\frac{\eta}{1-\eta}\right]^{1/2}
     - \sin^2 \pi g\;\left[\frac{1}{\eta(1-\eta)}\right]^{1/2}
\right) , \nonumber
\end{align}
revealing the conformal block structure in this simple example.
	
\subsubsection{Crossing symmetry}

The bulk two-point functions in the previous subsection can be
computed in another way, namely by first taking the OPE
of the two holomorphic fields and separately the OPE of the two
anti-holomorphic fields, and then applying the method of 
rotated images to the result of the anti-holomorphic OPE. The
chiral four-point function again reduces to a sum of chiral two-point 
functions, as indicated in figure~\ref{fig:crossing}, but since the 
rotated image operators are now in different 
representations of the $SU(2)$ algebra, the conformal blocks that
appear are not the same as before. Crossing symmetry is the statement
that the two methods give the same answer and in this subsection we verify 
that the method of rotated images preserves this symmetry. This provides
a non-trivial check on our prescription.
\psfrag{psi1}{$\psi_{j_1m_1}(z)$}
\psfrag{psi2}{$\psi_{j_2m_2}(w)$}
\EPSFIGURE{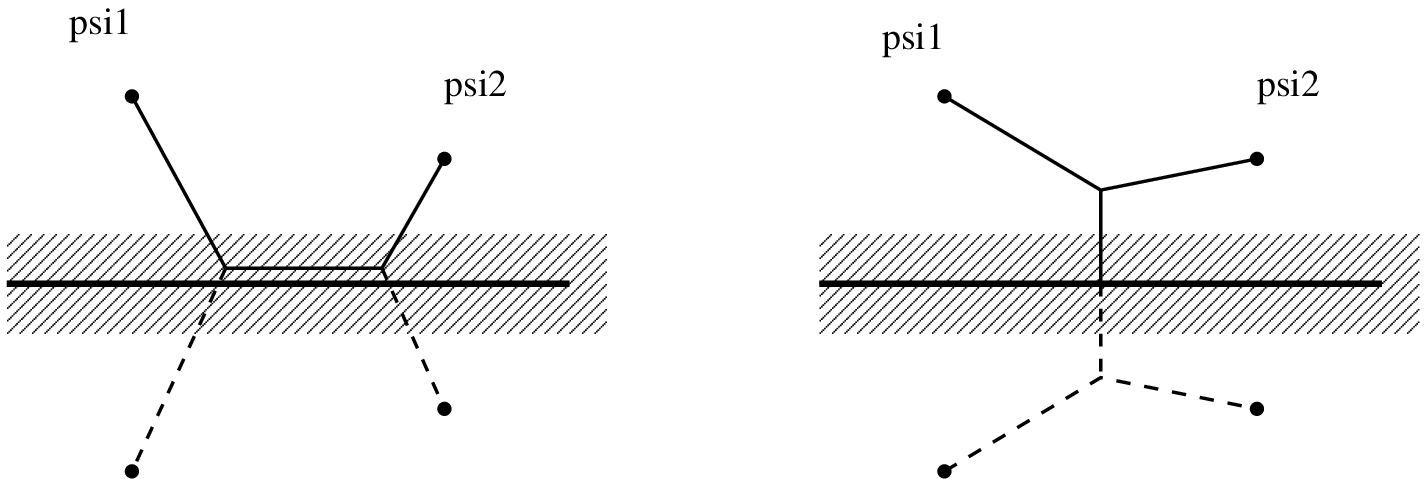, width=10cm}{
Crossing symmetry.
\label{fig:crossing}}

Crossing symmetry for 
$\langle
\psi_{j_1 m_1}(z)\psi_{\bar\jmath_1 \bar m_1}(z^*)
\psi_{j_2 m_2}(w)\psi_{\bar\jmath_2 \bar m_2}(w^*)
\rangle$
in the free theory can be expressed as 
\begin{align}
\label{freesym}
\sum_{J_p}
A^{J_p (m_1+\bar m_1)}_{j_1 m_1, \bar\jmath_1 \bar m_1}
&A^{J_p (m_2+\bar m_2)}_{j_2 m_2, \bar\jmath_2 \bar m_2}
\mathcal{F}^{(j_1 m_1)(\bar\jmath_1 \bar m_1)}_{(j_2 m_2)(\bar\jmath_2 \bar m_2)}
(J_p|\eta) 
\\ &= \sum_{J_q}
A^{J_q (m_1+m_2)}_{j_1 m_1, j_2 m_2}
A^{J_q (\bar m_1+\bar m_2)}_{\bar\jmath_1 \bar m_1, \bar\jmath_2 \bar m_2}
\mathcal{F}^{(j_1 m_1)(j_2 m_2)}_{(\bar\jmath_1 \bar m_1)(\bar\jmath_2 \bar m_2)}
(J_q|1-\eta).
\nonumber
\end{align}
The prefactor that accompanies the conformal blocks in \eqref{eq:genbulk2pt}
is the same in both channels and therefore cancels in the crossing relation.
With the boundary interaction turned on, we rotate 
$\psi_{\bar\jmath_1 \bar m_1}$ and $\psi_{\bar\jmath_2 \bar m_2}$ 
on the left hand side, but the product field 
$\psi_{J_qM_q}$, with $M_q= \bar m_1+\bar m_2$, on the right hand side.  
The requirement of crossing symmetry then becomes
\begin{align}
\label{eq:crossym}
\sum_{\bar m_1', \bar m_2'}
&\mathcal{D}^{\bar\jmath_1}_{\bar m_1,\bar m_1'}
 \mathcal{D}^{\bar\jmath_2}_{\bar m_2,\bar m_2'}
\sum_{J_p}
A^{J_p (m_1+\bar m_1')}_{j_1 m_1, \bar\jmath_1 \bar m_1'}
A^{J_p (m_2+\bar m_2')}_{j_2 m_2, \bar\jmath_2 \bar m_2'}
\mathcal{F}^{(j_1 m_1)(\bar\jmath_1 \bar m_1')}_{(j_2 m_2)(\bar\jmath_2 \bar
m_2')}(J_p|\eta)
\\ &= 
\sum_{J_q}
A^{J_q (m_1+m_2)}_{j_1 m_1, j_2 m_2}
A^{J_q (\bar m_1+\bar m_2)}_{\bar\jmath_1 \bar m_1, \bar\jmath_2 \bar m_2}
\mathcal{D}^{J_q}_{(\bar m_1+\bar m_2),(-m_1-m_2)}
\mathcal{F}^{(j_1 m_1)(j_2 m_2)}_{(\bar\jmath_1,-m_1)(\bar\jmath_2,-m_2)}
(J_q|1-\eta).
\nonumber
\end{align}
The conformal block 
$\mathcal{F}^{(j_1 m_1)(j_2 m_2)}_{(\bar\jmath_1 \bar m_1)(\bar\jmath_2 \bar m_2)}
(J_q|1-\eta)$
does not depend on the $m$'s individually, but only on the sums
$m_1+m_2$ and $\bar m_1+\bar m_2$, and due to momentum conservation the 
conformal block vanishes unless $m_1+m_2 = -(\bar m_1+\bar m_2)$. 
In the equation above we used $\bar m_1' = -m_1$ and $\bar m_2'=-m_2$ to 
express the last conformal block, but we could just as well have used other 
$\bar m$'s with the same sum.

To see that crossing symmetry is satisfied we start with the left hand side 
of equation (\ref{eq:crossym}), use the free crossing symmetry 
(\ref{freesym}) on the inner sum and
rewrite the product of the rotation coefficients using 
the relation \cite{Hamermesh:1962,Gaberdiel:2001xm}
\begin{align}
\mathcal{D}^{j_1}_{m_1,n_1}(g)\mathcal{D}^{j_2}_{m_2,n_2}(g)
= \sum_{J=\vert j_1-j_2 \vert}^{j_1+j_2}
C^{J (m_1+m_2)}_{j_1 m_1,j_2 m_2}
C^{J (n_1+n_2)}_{j_1 n_1,j_2 n_2}
\mathcal{D}^{J}_{(m_1+m_2),(n_1+n_2)}(g)
\end{align}
where the $C^{J M}_{j_1 m_1,j_2 m_2}$ are
Clebsch-Gordan coefficients. The left hand side of (\ref{eq:crossym})
becomes
\begin{align}
\label{lefthandside}
\sum_{\bar m_1', \bar m_2'}
\sum_{J} 
C^{J (\bar m_1+\bar m_2)}_{\bar\jmath_1 \bar m_1, \bar\jmath_2 \bar m_2}
&C^{J (\bar m_1'+\bar m_2')}_{\bar\jmath_1 \bar m_1', \bar\jmath_2 \bar m_2'}
\mathcal{D}^{J}_{(\bar m_1+\bar m_2),(\bar m_1'+\bar m_2')}
\\ &\times 
\sum_{J_q}
A^{J_q (m_1+m_2)}_{j_1 m_1, j_2 m_2}
A^{J_q (\bar m_1'+\bar m_2')}_{\bar\jmath_1 \bar m_1', \bar\jmath_2 \bar m_2'}
\mathcal{F}
^{(j_1 m_1) (j_2 m_2)}
_{(\bar\jmath_1 \bar m_1')(\bar\jmath_2\bar m_2')}
(J_q|1-\eta).
\nonumber
\end{align}
Now we recall from equation (\ref{eq:AfC}) that 
$A^{J M}_{j m, j' m'} = f(j,j',J) C^{J M}_{jm, j'm'}$
and therefore 
\begin{align}
A^{J_q (\bar m_1'+\bar m_2')}_{\bar\jmath_1 \bar m_1', \bar\jmath_2 \bar m_2'}
=
A^{J (\bar m_1+\bar m_2)}_{\bar\jmath_1 \bar m_1, \bar\jmath_2 \bar m_2} \,
\frac{C^{J_q (\bar m_1'+\bar m_2')}_{\bar\jmath_1 \bar m_1', \bar\jmath_2 \bar m_2'}%
    }{C^{J (\bar m_1+\bar m_2)}_{\bar\jmath_1 \bar m_1, \bar\jmath_2 \bar m_2}}
\, \frac{f(\bar\jmath_1, \bar\jmath_2, J_q)}{f(\bar\jmath_1, \bar\jmath_2, J)}.
\end{align}
Making this substitution in (\ref{lefthandside}), the 
$C^{J(\bar m_1+\bar m_2)}_{\bar\jmath_1 \bar m_1, \bar\jmath_2 \bar m_2}$ 
cancels and we get
\begin{align}
\sum_{J, J_q} 
\sum_{\bar m_1', \bar m_2'}'
C^{J_q (\bar m_1'+\bar m_2')}_{\bar\jmath_1 \bar m_1', \bar\jmath_2 \bar m_2'}
&C^{J (\bar m_1'+\bar m_2')}_{\bar\jmath_1 \bar m_1', \bar\jmath_2 \bar m_2'}
\mathcal{D}^{J}_{(\bar m_1+\bar m_2),(-m_1-m_2)}
f(J_q)/f(J) \nonumber
\\ &\times
A^{J_q (m_1+m_2)}_{j_1 m_1, j_2 m_2}
A^{J (\bar m_1+\bar m_2)}_{\bar\jmath_1 \bar m_1, \bar\jmath_2 \bar m_2}
\mathcal{F}^{(j_1 m_1)(j_2 m_2)}_{(\bar\jmath_1, -m_1)(\bar\jmath_2, -m_2)}
(J_q\vert 1-\eta)
\end{align}
where we have replaced $\bar m_1'+\bar m_2' \to -m_1-m_2$ in the rotation 
coefficient and $\bar m_1' \to -m_1$ and $\bar m_2' \to -m_2$ in the conformal 
block. The prime  on the sum over $\bar m_1'$ and $\bar m_2'$ means
that we restrict the sum to $\bar m_1'+\bar m_2' = -m_1-m_2$.  This can be
done because the conformal block vanishes for other values of 
$\bar m_1'+\bar m_2'$.
The sum over the Clebsch-Gordan coefficients is
\begin{align}
\sum_{\bar m_1', \bar m_2'}'
C^{J_q (\bar m_1'+\bar m_2')}_{\bar\jmath_1 \bar m_1', \bar\jmath_2 \bar m_2'}
C^{J (\bar m_1'+\bar m_2')}_{\bar\jmath_1 \bar m_1', \bar\jmath_2 \bar m_2'}
&=
\sum_{\bar m_1', \bar m_2'}
\langle J_q M \vert \bar\jmath_1 \bar m_1', \bar\jmath_2 \bar m_2' \rangle
\langle \bar\jmath_1 \bar m_1', \bar\jmath_2 \bar m_2' \vert J M \rangle
\\ &= \langle J_q M \vert J M \rangle
= \delta(J_q-J)
\end{align}
where $M = -m_1-m_2$ and we have used the completeness relation for 
$\vert \bar\jmath_1 \bar m_1', \bar\jmath_2 \bar m_2' \rangle$. 
Finally, we carry out the sum over $J$ to arrive at
\begin{align}
\sum_{J_q} 
A^{J_q (m_1+m_2)}_{j_1 m_1, j_2 m_2}
A^{J_q (\bar m_1+\bar m_2)}_{\bar\jmath_1 \bar m_1, \bar\jmath_2 \bar m_2}
\mathcal{D}^{J_q}_{(\bar m_1+\bar m_2),(-m_1-m_2)}
\mathcal{F}^{(j_1 m_1)(j_2 m_2)}_{(\bar\jmath_1,-m_1)(\bar\jmath_2,-m_2)}
(J_q\vert 1-\eta)
\end{align}
which is exactly the right hand side of equation (\ref{eq:crossym}).

\section{Boundary fields}
\label{boundaryfields}

When a bulk field approaches the boundary at $z=\bar z$, new divergences
appear that are not removed by the bulk normal ordering.  This is a 
general feature of conformal field theories with boundaries and signals
the presence of so-called {\it boundary fields} in the 
theory \cite{Diehl:1981}.
A bulk field approaching the boundary has an operator product expansion,
\begin{equation}
\label{bbope}
\Psi_{h,\bar h}(z, \bar z) =
\sum_i \frac{B^i_{h,\bar h}}{(z-\bar z)^{h+\bar h - \Delta_i}} 
\, \Psi_i^B(x) \, ,
\end{equation}
where $x = (z+\bar z)/2\,$.  The $\Psi_i^B(x)$ are boundary fields 
with boundary scaling dimensions $\Delta_i$ and the 
$B^i_{h,\bar h}$ are called bulk-to-boundary operator product coefficients.  

In addition to the bulk-to-boundary OPE, the boundary fields form an 
operator product algebra amongst themselves,
\begin{equation}
\label{boundaryope}
\Psi_i^B(x) \, \Psi_j^B(x') = \sum_k 
\frac{C_{ijk}}{(x-x')^{\Delta_i+\Delta_j-\Delta_k}} 
\, \Psi_k^B(x')\,.
\end{equation}
The boundary OPE coefficients $C_{ijk}$ and the bulk-to-boundary
OPE coefficients $B^i_{h,\bar h}$, along with the boundary scaling
dimensions $\Delta_i$, are characteristic data of a given boundary
conformal field theory.

In radial quantization boundary operators correspond to open string
states, defined on a semi-circle enclosing the insertion point on the
boundary.  The neighborhood of the insertion point can be mapped by a
conformal transformation to the strip, as shown in 
figure~\ref{fig:stateopB}.  
\psfrag{a}{\ }
\psfrag{b}{\ }
\EPSFIGURE{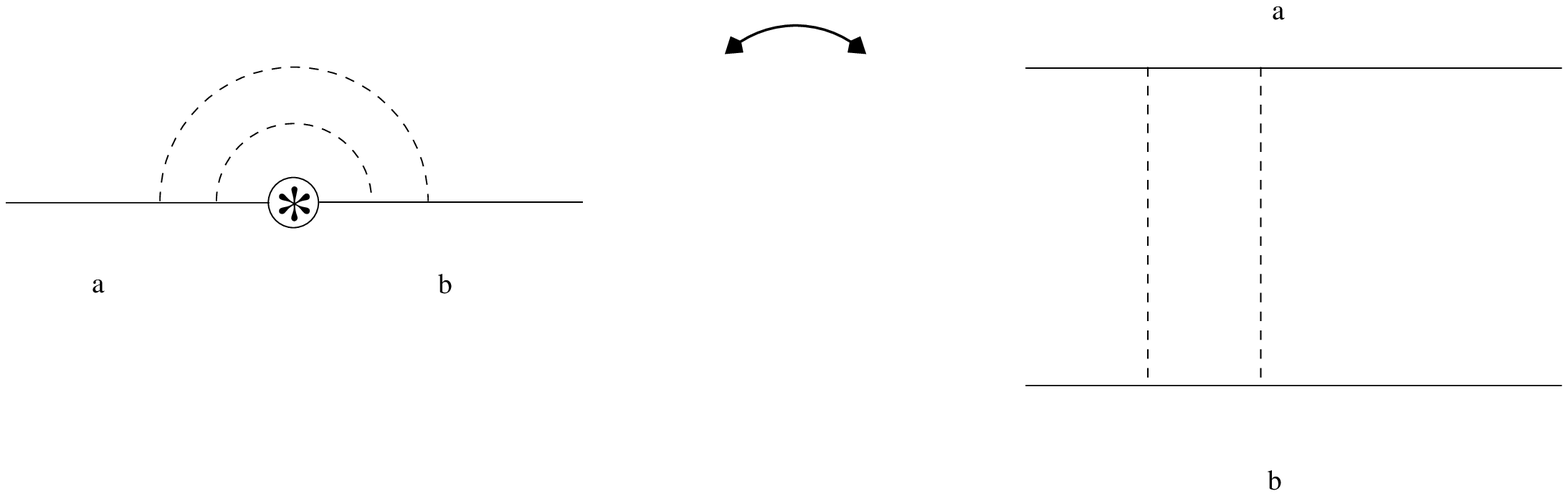, width=6cm}{
Correspondence between boundary operators and open string states via 
map between the halfplane and infinite strip.
\label{fig:stateopB}}
The boundary scaling dimension $\Delta_i$ of a boundary operator 
$\Psi_i^B(x)$ is given by the energy eigenvalue of the corresponding open 
string state, where the open-string Hamiltonian is the $L_0$ generator of 
the Virasoro algebra that is preserved by the conformally invariant boundary 
conditions. With this in mind, we will examine the open string spectrum of 
the interacting $c=1$ model defined by (\ref{action}) in detail in
section~\ref{openspectrum}, but first let us look at how all of this 
works in the free theory. 

\subsection{Free theory}

In this case we can use the doubling trick to replace the right-moving
part of the bulk operator in (\ref{bbope}) by its left-moving image,
\begin{equation}
\Psi_{h,\bar h}(z, \bar z) \rightarrow  \psi_h(z)\psi_{\bar h}(z^*) \,,
\end{equation}
and view the approach to the boundary as two left-moving fields
moving towards the real axis from opposite directions.
We can then work out the operator product using free-field Wick contractions 
and the holomorphic propagator (\ref{freeprop}).  Consider a non-compact
boson and let a bulk primary field at generic momentum approach the 
boundary. The chiral OPE between its left-moving part and the left-moving
image of the right-moving part is 
\begin{eqnarray}
e^{ik\phi(z)} e^{ik \phi(z^*)} &=& 
  \frac{e^{2ik \phi(\frac{z+z^*}{2})}}{(z-z^*)^{-k^2}} + \cdots \nonumber \\
  &=& \frac{e^{ik \Phi(x)}}{(z-z^*)^{-k^2}} + \cdots
\label{freeope}
\end{eqnarray}
where $x=(z+z^*)/2$ and the boundary scalar field $\Phi(x)$ is related to 
the left-moving field as before by $\Phi(x)=2 \phi(z)\vert_{z=z^*}$.  
The free-field propagator on the boundary is 
\begin{equation}
\label{boundaryfreeprop}
\langle \Phi(x) \Phi(x')\rangle = -4 \log{\vert x-x'\vert} \,,
\end{equation}
and thus the boundary scaling dimension of $\Psi^B_k(x)=e^{ik \Phi(x)}$ is
$\Delta_k=2 k^2$.  Since the bulk scaling dimension of $e^{ik\phi(z)}$ is 
$h=k^2/2$, we see that (\ref{freeope}) is consistent with the
general form (\ref{bbope}) and we can read off 
$B^k_{h,\bar h}=1$ for the bulk-to-boundary OPE coefficient.
The boundary OPE for primary fields of the form $e^{ik \Phi(x)}$ is 
given by 
\begin{equation}
\label{boundaryfreeope}
e^{ik_1\Phi(x)} e^{ik_2\Phi(x')} 
= \frac{e^{i(k_1+k_2)\Phi(x')}}{\vert x-x'\vert^{4k_1k_2}} +\cdots
\end{equation}
This has the form of \eqref{boundaryope} with 
$C_{k_1,k_2,k_3} = \delta_{k_1+k_2+k_3,0}$.

When a bulk discrete field approaches the boundary we get 
\begin{equation}
\label{eq:B2B0}
\psi_{jm}(z) \psi_{\bar\jmath \bar m}(z^*) =
\sum_{J=|j-\bar\jmath|}^{j+\bar\jmath}
\sum_{M=-J}^J
\frac{B^{JM}_{jm,\bar\jmath \bar m}}
     {(z-z^*)^{j^2 +\bar\jmath^2-J^2}}
     \Psi_{JM}(x)
+ \cdots
\end{equation}
where the $\Psi_{JM}(x)$ are primary fields on the boundary, 
which we will refer to as {\it discrete boundary fields}.
The discrete boundary fields come in $SU(2)$ multiplets just as the
chiral discrete fields, but since $j-\bar\jmath\in \mathbb{Z}$ 
only integer values of $J$ are allowed on the boundary.  
The currents of the boundary $SU(2)$ algebra are given by
\begin{equation}
\label{boundarycurrents}
J_\pm = e^{\pm i \Phi /\sqrt{2}}  
\quad \textrm{and} \quad J_3 = \frac{i}{2\sqrt{2}}\frac{d\Phi}{dx}.
\end{equation}
The discrete boundary fields can be easily obtained as follows. Work in
the theory at the self-dual radius and let the purely holomorphic bulk 
discrete primary field 
$\Psi_{jm,00}(z,\bar z) = \psi_{jm}(z)\bar\psi_{00}(\bar z)$
approach the boundary. Since $\bar\psi_{00}(\bar z)$ is the unit 
operator the bulk-to-boundary OPE only contains one primary on the right 
hand side and we find that
\begin{equation}
\label{bfields}
\Psi_{JM}(x)= \psi_{JM}(z) \vert_{z=x} \,.
\end{equation}
The discrete boundary field $\Psi_{J M}$ is thus made from 
$e^{iM\,\Phi /\sqrt{2}}$ times a polynomial in derivatives of $\Phi$
that is obtained from the one that appears in $\psi_{JM}(z)$ by the
replacement $\partial\Phi \rightarrow \Phi'/2$. The boundary 
scaling dimension of $\Psi_{J M}$ is $J^2$.
This construction of discrete boundary fields is in the theory at the 
self-dual radius but the resulting $\Psi_{J M}$ boundary fields also 
exist at other radii. They form a closed algebra under the boundary OPE
\eqref{boundaryope}, with contractions carried out using the
free boundary propagator \eqref{boundaryfreeprop}.

It follows from the above construction that in the free theory the 
bulk-to-boundary operator product coefficients 
$B^{JM}_{jm,\bar\jmath \bar m}$ for discrete primary fields are given 
precisely by the chiral OPE coefficients $A^{JM}_{jm,\bar\jmath \bar m}$ 
in \eqref{eq:chiralope} for integer $J$.
Finally, we note that the field $\Psi_{J M}$ could equally well have 
been obtained from the anti-holomorphic bulk field $\Psi_{00,JM}$.

\subsection{Interacting theory}
\label{sec:intboundaryfields}

The boundary fields are modified when the interaction 
\eqref{bpotential} is turned on.
The boundary OPE can no longer be evaluated using the free field boundary
propagator \eqref{boundaryfreeprop} and the boundary primary field 
$\Psi_p(x)$ carrying generic momentum $p$ no longer has the simple form 
$e^{ik\Phi(x)}$.  This is evident from the change 
in the scaling dimensions of boundary primary operators at generic 
momentum. The boundary scaling dimensions correspond to $L_0$ 
eigenvalues of the corresponding open string states and the open string 
spectrum undergoes a non-trivial ``flow'' as a function of the boundary 
coupling $g$. Bands are formed with gaps between them that grow wider 
with increasing coupling strength \cite{Polchinski:1994my}. We discuss 
the open string spectrum in detail in section~\ref{openspectrum}. 
The remaining data that defines the interacting boundary conformal 
field theory, {\it i.e.} the bulk-to-boundary OPE and the boundary OPE, 
will also depend on coupling $g$. Conformal invariance places restrictions 
on low-order correlation functions of boundary operators, but beyond that 
little is known. 

At the momenta carried by the boundary discrete fields the situation
is simplified. As we will see in section~\ref{openspectrum}, the 
dimension of discrete boundary fields $\Psi_{J M}$ does not flow when 
the boundary interaction is turned on but remains at $J^2$. 
This enables us to write a discrete boundary field in the 
interacting theory $\Psi^g_{J M}$ as a linear combination of the 
free boundary fields in the $SU(2)$ representation with the same 
$J$ value,
\begin{equation} 
\label{eq:2a}
\Psi^g_{J M}(x)=\sum_{M'=-J}^J h^J_{MM'}(g) \Psi^0_{J M'}(x) \,.
\end{equation}
The bulk-to-boundary OPE for discrete fields at non-zero boundary coupling
is given by
\begin{equation}
\Psi_{j m,\bar\jmath \bar m}(z,\bar z) 
= \sum_{J=\vert j- \bar\jmath \vert}^{j+\bar\jmath} \sum_{M=-J}^{J} 
\frac{B^{JM}_{jm,\bar\jmath \bar m}(g)}
{(z-z^*)^{j^2+\bar\jmath^2-J^2}}\Psi^g_{J M}(x) + \ldots 
\label{interactingb2bope}
\end{equation} 
The bulk discrete field is unaffected by the boundary physics but 
in general both the bulk-to-boundary OPE coefficients and the boundary 
discrete fields may be expected to depend on $g$.  

Now we apply the method of rotated images to the bulk primary field 
on the left hand side of \eqref{interactingb2bope} and then use
the free theory bulk-to-boundary OPE \eqref{eq:B2B0} to obtain,
\begin{equation}
\psi_{jm}(z) \tilde\psi_{\bar\jmath\bar m}(z^*)
=\sum_{\bar m'=-\bar\jmath}^{\bar\jmath} 
\sum_{J=\vert j- \bar\jmath \vert}^{j+\bar\jmath} \sum_{M=-J}^{J}
\mathcal{D}^{\bar\jmath}_{\bar m,\bar m'}(g) \,
\frac{B^{JM}_{jm,\bar\jmath \bar m'}(0)}
{(z-z^*)^{j^2+\bar\jmath^2-J^2}}\Psi^0_{J M}(x) + \ldots
\label{rotatedb2bope}
\end{equation}
The right hand side is to equal that of \eqref{interactingb2bope} and 
thus, by inserting the expansion (\ref{eq:2a}) for $\Psi^g_{J M}(x)$, 
we obtain algebraic equations that relate the bulk-to-boundary OPE 
coefficients $B^{JM}_{jm,\bar\jmath \bar m}(g)$ and the expansion 
coefficients $h^J_{MM'}(g)$,
\begin{equation}
\sum_{M'=-J}^{J}
B^{JM'}_{jm,\bar\jmath \bar m}(g) \,
h^J_{M'M}(g) 
=\sum_{\bar m'=-\bar\jmath}^{\bar\jmath}
\mathcal{D}^{\bar\jmath}_{\bar m,\bar m'}(g) \,
B^{JM}_{jm,\bar\jmath \bar m'}(0) \,.
\label{algebraicrelation}
\end{equation}
Unfortunately, there are not enough equations to determine all the 
coefficients, so we need further input. 

It seems reasonable to define the discrete boundary field 
$\Psi^g_{J M}(x)$ in the interacting theory in the same fashion as in
the free theory, {\it i.e.} as the boundary primary field that arises
when we let the purely holomorphic bulk field $\Psi_{JM,00}$ approach
the boundary. If we then apply the method of rotated images, we find
that the global $SU(2)$ rotation acts trivially on the unit operator
$\psi_{00}(z^*)$ and the discrete boundary field is actually the same
as in the free theory,
\begin{equation}
h^J_{MM'}(g) = \delta_{M,M'} \,.
\label{trivialh}
\end{equation}
It follows that all boundary correlators of discrete fields will be
independent of the boundary coupling $g$. This is a surprisingly strong
result and certainly only applies to the discrete boundary fields.
At generic momentum the boundary correlators depend on $g$ in a 
non-trivial way.

When applying the method of rotated images to $\Psi_{JM,00}$ we deformed
the integration contours of the boundary currents into the lower 
half-plane where they encountered only $\psi_{00}(z^*)$. We could equally 
well have moved the contours into the upper half-plane, leading to 
\begin{equation}
h^J_{MM'} = \mathcal{D}^J_{MM'}(-g) \,.
\label{rotatedh}
\end{equation}
In this case the algebraic relation \eqref{algebraicrelation} between 
$h^J_{MM'}(g)$ and the bulk-to-boundary OPE coefficients is modified,
\begin{equation}
\sum_{M'=-J}^{J}
B^{JM'}_{jm,\bar\jmath \bar m}(g) \,
h^J_{M'M}(g) 
=\sum_{m'=-j}^j
\mathcal{D}^j_{m,m'}(-g) \,
B^{JM}_{jm',\bar\jmath \bar m}(0) \,,
\label{newalgebraicrelation}
\end{equation}
and the OPE coefficients that solve them are not the same as for
\eqref{algebraicrelation}.
The new prescription nevertheless leads to the same boundary correlators 
as before.  This is because \eqref{rotatedh} amounts to the same 
$g$-dependent $SU(2)$ rotation acting on all the discrete boundary fields 
of the free theory, which leaves their correlation functions unchanged,
\begin{align}
\label{boundaryamplitudes}
&\langle\Psi^g_{J_1M_1}(x_1)\ldots\Psi^g_{J_nM_n}(x_n)\rangle\nonumber\\
&=\sum_{M'_i=-J_i}^{J_i}
\mathcal{D}^{J_1}_{M_1M'_1}(g)\ldots\mathcal{D}^{J_n}_{M_nM'_n}(g)
\langle\Psi^0_{J_1M'_1}(x_1)\ldots\Psi^0_{J_nM'_n}(x_n)\rangle \\
&=\langle\Psi^0_{J_1M_1}(x_1)\ldots\Psi^0_{J_nM_n}(x_n)\rangle\,.\nonumber
\end{align}

We could also have defined $\Psi_{JM}(x)$ as the boundary
field obtained when we let the anti-holomorphic bulk field
$\Psi_{00,JM}$ approach the boundary. In the free theory this
definition is equivalent to the one based on a purely holomorphic
bulk field and for correlation functions involving discrete 
boundary fields this remains true in the interacting
theory, they are independent of $g$ with either definition. As we 
will see in section~\ref{sec:mixedcorrel} below, however, mixed 
amplitudes containing both bulk and boundary fields will in general 
depend on which definition we use. 
 
\subsection{Boundary condition changing fields}

We can also consider more general boundary fields which change the
boundary conditions where they are inserted into correlation
functions. The map from the upper half-plane to the strip reveals 
that the corresponding open string states are subject to different 
boundary conditions at the two endpoints, as 
indicated in figure~\ref{fig:bcchanging}.
\psfrag{a}{$\alpha$}
\psfrag{b}{$\beta$}
\EPSFIGURE{open.eps, width=6cm}{
Boundary condition changing fields correspond to open strings with different 
boundary conditions at the two endpoints.
\label{fig:bcchanging}}

In the theory at hand, boundary conditions are labelled by the boundary 
coupling $g$ which multiplies the periodic boundary potential in
(\ref{action}).  A boundary condition changing operator
changes $g$ to $g'$ at the insertion point on the boundary and thus
corresponds to open string endpoints interacting with boundary potentials 
of different strength $g$ and $g'$. We obtain the spectrum of such 
strings in section~\ref{twogs} below.

\section{The open string spectrum}
\label{openspectrum}

The scaling dimensions of boundary operators in a given boundary conformal 
field theory are given by the $L_0$ eigenvalues of the corresponding open 
string states. In this section we will mostly work with a non-compact
boson, for which the open string spectrum is continuous.  The discrete 
spectra obtained at finite boson radii are all included in the continuous 
spectrum and are obtained by retaining only those eigenvalues that 
correspond to allowed momenta in each case.  

The information we are after is for example contained in the one-loop 
partition function in the open string channel,
\begin{equation}
\mathcal{Z} = {\rm Tr}\,[\exp(-\beta H_\mathrm{open})] \,.
\end{equation}
For open strings with Neumann boundary conditions at both ends 
a standard calculation gives
\begin{equation}
\mathcal{Z}_\mathrm{free} = \frac{1}{\eta(\omega)} \int_{-\infty}^\infty
\frac{dp}{2\pi}\, \omega^{p^2/2} \,,
\end{equation}
where $\eta(\omega)=\omega^{1/24}\prod_{n=1}^\infty (1-\omega^n)$ 
is the Dedekind eta-function with $\omega=e^{-\pi\beta/\ell}$. Here 
$\ell$ is the parameter length of the open string.
The $p$ integral can be interpreted as a sum over Virasoro 
representations obtained from primary fields of the form
$e^{ip\Phi(x)/\sqrt{2}}$.

\subsection{Spectral flow and string band structure}

The open string spectrum is modified in the presence of the periodic 
boundary potential (\ref{bpotential}). The interacting theory exhibits a
non-trivial spectral flow that depends on the strength of the boundary
coupling $g$. The system can be re-expressed in terms of free 
fermions as shown in \cite{Polchinski:1994my}.  The $SU(2)$ currents are 
bi-linear in the fermions and the boundary interaction may be viewed as 
a localized mass term.  The open string spectrum is then found by 
solving a straightforward eigenvalue problem for the fermions.  We will 
not repeat the construction here but simply quote the result.  In 
\cite{Polchinski:1994my} both string endpoints were taken to 
interact with the same boundary potential, $g=g'$.  In that case  
the partition function may be written 
\begin{equation}
\label{partfcn}
\mathcal{Z} = \frac{\sqrt{2}}{\eta(\omega)} \int_{-1/2}^{1/2} \frac{dk}{2\pi}  
      \sum_{m=-\infty}^{\infty} \omega^{(\lambda + m)^2} \,,
\end{equation}
where $\lambda$ is related to the target space momentum $p=\sqrt{2}k$ by
\begin{equation}
\label{eq:spectrum}
\sin \pi \lambda = \cos \pi |g| \sin \pi k  \,.  
\end{equation}
The value of $\lambda$ lies within the range 
$-\half+\vert g\vert \leq \lambda \leq \half -\vert g\vert$ and the 
energy eigenvalues of the highest-weight open string states are given by 
$\Delta = \left(\lambda(k) + m\right)^2$, with $m\in \mathbb{Z}$.

The spectrum is shown in the left-most graph in figure~\ref{fig:genspectrum}.
It has split into bands with forbidden gaps in energy in between them.  
Note that the energy eigenvalues are invariant under $k \to k + n$ for any 
integer $n$. This is due to the fact that the boundary potential breaks 
translation invariance in the target space to a discrete subgroup.  In the 
interacting system target space momentum $p$ is therefore only conserved 
mod $\sqrt{2}$ in our units, and a shift of $k$ by an integer can always be
used to bring the momentum of a given operator into the so-called first
Brillouin zone, $-\frac{1}{\sqrt{2}}\leq p \leq \frac{1}{\sqrt{2}}$.  
The spectra in figure~~\ref{fig:genspectrum} are displayed in an extended 
zone scheme, with the periodicity in momentum appearing explicitly.
The boundary fields $V_{n,p}(x)$ that correspond to these open string states 
carry two labels. The integer $n\geq 0$ denotes the band, to which the 
state belongs, and $p$ is the momentum within the first Brillouin zone.

The allowed values of $\lambda + m$ in \eqref{partfcn}
consist of bands of width $1-2|g|$ centered at every integer, with gaps
of width $2|g|$ between. 
When $g=0$ we get $\lambda +m = k$ and the gaps disappear.  This
corresponds to free propagation of open strings with Neumann boundary 
conditions at both ends.  Equation (\ref{eq:spectrum}) gives 
$\sin\pi\lambda = 0$ when $\vert g\vert = 1/2$ and $\lambda$ must 
then be an integer.  This is the tight binding limit, the
bands have zero width and we get Dirichlet boundary conditions confining
each string end to sit at one of the minima 
of the boundary potential.\footnote{By looking at the boundary potential 
in (\ref{action}) one might have expected tight binding to occur for 
$\vert g\vert\rightarrow\infty$ rather than $\vert g\vert\rightarrow 1/2$.  
As discussed in \cite{Callan:1994ub,Kogetsu:2004te}, this is a 
renormalization effect.  There are divergences in the model and the formula 
(\ref{eq:spectrum}) for the spectrum implicitly refers to a particular 
regularization and subtraction procedure which redefines the coupling 
constant according to \eqref{renormg}.} 
This also has an interpretation as a sequence of D-particles, evenly
spaced along the $\Phi$ direction in target space.

In conventional condensed matter systems, band gaps get vanishingly small 
as the energy becomes large compared to the characteristic scale set by 
the potential.  As a consequence of conformal invariance, the stringy 
bands discussed here behave rather differently.  Both the individual bands 
and the gaps between them grow wider with increasing energy.

It follows from \eqref{eq:spectrum} that the scaling dimension 
of a discrete boundary field $\Psi_{JM}(x)$ does not undergo 
spectral flow. These fields carry momenta that correspond to 
$k\in\mathbb{Z}$, for which the right hand side of \eqref{eq:spectrum}
is zero for any value of $g$. This observation plays a key role
when we discuss the discrete boundary operators in the interacting 
theory.

When the boson is compactified we have to restrict to allowed momentum
values and the energy bands are discretized accordingly.
For a boson at the self-dual radius, we only have the discrete fields. 
Since their scaling dimensions are unaffected, the full open string 
spectrum is in fact independent of the boundary interaction in this case, 
as was first observed in \cite{Callan:1994ub}.

It is interesting to note that energy bands also appear in the open string
channel of the boundary conformal field theory that describes a free boson 
compactified on a circle, when the radius of the circle is irrational
\cite{Friedan:1999,Janik:2001hb}, {\it i.e.} not a rational number 
times the self-dual radius $R_\mathrm{sd}=\sqrt{2}$.  Like the present system, 
those theories admit a one-parameter family of boundary states that 
interpolate between Neumann and Dirichlet boundary conditions, but it is 
unclear to us how deep the parallels between the systems run.

\subsection{Open strings coupled to two different boundaries}
\label{twogs}

The band spectrum found in \cite{Polchinski:1994my} is
a special case of a more general structure.  When we allow for 
boundary condition changing operators we have to consider also
open strings where the boundary coupling takes different 
values $g_1$ and $g_2$ at the endpoints.

The fermion eigenvalue problem solved in \cite{Polchinski:1994my} can 
easily be extended to cover this case also.\footnote{The spectrum can
also be obtained by evaluating the appropriate one-loop partition 
function in the closed-string channel. This calculation is presented
in appendix~\ref{app:closedchannel}.} 
The only modification is to 
change $g$ to $g_1$ in one of the boundary mass terms for the worldsheet
fermions in equation (29) of that paper and $g$ to $g_2$ in the other one.
The relevant eigenvalue equation becomes
$$
e^{-iN_1}e^{iN_2}e^{iN_3} \Psi = \Lambda \Psi
$$
where
\begin{align}
\label{NNN}
N_1 = \pi \left(
\begin{array}{cc} 0& wg_1 \\ \bar w\bar g_1 & 0\end{array}
\right), \quad
N_2 = \pi \left(
\begin{array}{cc} 0& g_2 \\ \bar g_2 & 0\end{array}
\right), \quad
N_3 = 2\pi k \sigma^3
\end{align}
and $w = e^{-2\pi i k}$.
For simplicity we assume that $g_1,g_2 \in \mathbb{R}$, with $g_1\geq g_2$,
in the following.
We notice that
$$
\det{e^{-iN_1}e^{iN_2}e^{iN_3}}
= \exp \left(\tr e^{-iN_1} \tr e^{iN_2} \tr e^{iN_3} \right) = 1
$$
and find that
\begin{equation}
\label{trace}
\tr\left( e^{-iN_1}e^{iN_2}e^{iN_3} \right) =
2\cos(\pi g_1) \cos (\pi g_2) \cos (2\pi k)
+ 2 \sin (\pi g_1) \sin(\pi g_2) .
\end{equation}
The trace is real and lies between $-2$ and $2$.
The characteristic equation has a negative
discriminant and real coefficients so the
eigenvalues are complex conjugates of one another,
$\Lambda_1 = \bar \Lambda_2$. Furthermore,
$\Lambda_1\Lambda_2 = 1$ so we can write
$\Lambda_{1,2} = e^{\pm 2\pi i \lambda}$. The
trace is equal to the sum of the eigenvalues
and the equation for $\lambda$ becomes
$$
2\cos(2\pi\lambda)
= 2\cos(\pi g_1) \cos(\pi g_2) \cos(2\pi k)
+2 \sin(\pi g_1) \sin(\pi g_2)
$$
which can be rewritten as
\begin{equation}
\label{eq:generalspectrum}
\sin^2(\pi\lambda)
= \sin^2(\pi g_-)\cos^2(\pi k)
+ \cos^2(\pi g_+)\sin^2(\pi k)
\end{equation}
with $g_\pm = \half(g_1 \pm g_2)$.
Clearly this reduces to the previous result (\ref{eq:spectrum})
as $g_1 \to g_2$ but when $g_1\ne g_2$ there are important new features.  
In particular, there are additional gaps in the spectrum as shown in 
figure~\ref{fig:genspectrum}. 
\psfrag{En}{\footnotesize $L_0$}
\psfrag{k}{\footnotesize $k$}
\psfrag{1}{\footnotesize 1}
\psfrag{4}{\footnotesize 4}
\EPSFIGURE[h]{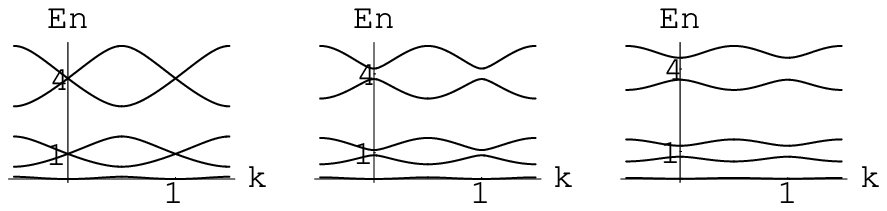, width=\linewidth}{
The first few bands of the energy spectrum for different values 
of the coupling constants. One is held fixed at $g_2=0.2$ but the other
takes the values a) $g_1=0.2$, b) $g_1=0.3$, c) $g_1=0.4$.
\label{fig:genspectrum}}

By varying either $g_1$ or $g_2$ from zero to one-half we effectively 
interpolate between Neumann and Dirichlet boundary conditions at that 
end.  If one string end is subject to a Dirichlet boundary condition, 
say $g_1=1/2$, the spectrum is discrete, with eigenvalues $(m+1/2 \pm g_2)^2$, 
and if both ends have Neumann boundary conditions, $g_1=g_2=0$, the spectrum 
varies continuously over all positive real numbers.  For all other values 
of $g_1$ and $g_2$ the spectrum exhibits band structure.  It starts with a 
gap from zero energy to $(g_1 - g_2)^2/4$, followed by a band of width 
$(\frac{1}{2} - g_1)(\frac{1}{2} - g_2)$.
After that two types of gaps alternate.  For each positive integer $m$, 
there are gaps of width $2m(g_1-g_2)$ and $(2m+1)(g_1+g_2)$ in $\lambda$,
whereas the intervening bands have widths 
$(\frac{1}{2} -g_1) (2m - (\frac{1}{2} -g_2))$ 
and $(\frac{1}{2} -g_1) (2m + (\frac{1}{2} -g_2))$.
The limit $g_1=g_2$ is very special because then half of all the gaps close.

At finite boson radius the general band structure \eqref{eq:generalspectrum}
associated with boundary condition changing fields is rendered discrete by 
the restriction to allowed momenta. At the self-dual radius the result is a 
particularly simple set of eigenvalues $(n\pm g_-)^2$, with $n\in\mathbb{Z}$.

\subsection{Open string spectrum in the half-brane theory}
\label{halfbranespectrum}

By setting $\bar g_1 = \bar g_2 = 0$ in equation (\ref{NNN})
we can obtain the open string spectrum for the `half-brane'
theory (\ref{complexaction})
\begin{equation}
  S = \frac{1}{4\pi} \int d^2 z\: \partial \Phi \bar \partial \Phi
    - \frac{1}{2} g_0 \int d\tau e^{i\Phi(\tau)/\sqrt{2}} \,.
\end{equation}
The trace in (\ref{trace}) reduces to $2\cos(2\pi k)$ and we 
immediately find that $\lambda =k$. The spectrum is therefore
unaffected by the half-brane boundary interaction.  This result was 
previously found in \cite{Gaberdiel:2004na} by a different method.

\section{Boundary correlation functions}
\label{boundarycorrelators}

In section~\ref{sec:intboundaryfields} we argued that correlation 
functions involving discrete boundary fields are not affected by the
boundary interaction. This only applies for discrete fields and 
correlators of more general boundary fields will depend on the
boundary coupling in a non-trivial way.

\subsection{Boundary fields carrying generic momentum}

We can anticipate the structure of low-order boundary 
amplitudes from conformal symmetry. The boundary primary field at
generic momentum is $V_{n,p}(x)$, where $n\geq 0$ labels the band that 
the corresponding open string state lies in and $p$ is a momentum 
in the first Brillouin zone, 
$-\frac{1}{\sqrt{2}}\leq p \leq \frac{1}{\sqrt{2}}$.
For two- and three-point functions one finds
\begin{align}
\langle V_i(x_i) V_j(x_j)\rangle
=& \frac{G_{ij}}{|x_i-x_j|^{2\Delta_i}} \,, \\
\langle V_i(x_i) V_j(x_j) V_k(x_k)\rangle
=& \frac{C_{ijk}}{
|x_i-x_j|^{\Delta_i + \Delta_j - \Delta_k}
|x_j-x_k|^{\Delta_j + \Delta_k - \Delta_i}
|x_k-x_i|^{\Delta_k + \Delta_i - \Delta_j}} \,,\nonumber
\end{align}
where $V_i$ is short for $V_{n_i,p_i}$ and $\Delta_i=(n_i+\lambda(p_i))^2$ 
is the scaling dimension obtained from \eqref{eq:spectrum}, or 
\eqref{eq:generalspectrum} in the case of boundary condition changing
fields.  It remains to determine the $g$-dependence of the coefficients 
$G_{ij}$ and $C_{ijk}$ in the interacting theory. Conformal 
invariance requires the boundary scaling dimensions of $V_i$ and
$V_j$ in the boundary two-point function to be equal. 
We can normalize the boundary fields in such a way 
that $G_{ij}=\delta_{n_i,n_j}$ if $p_i+p_j =0$ and $G_{ij}=0$ 
otherwise. This means that, once we have determined the open string
spectrum, we know the two-point functions of all boundary fields in 
the interacting theory. This way the coupling dependence of $G_{ij}$
is shifted into the bulk-to-boundary OPE coefficients for fields at
generic momentum. These OPE coefficients can in principle be obtained
from bulk two-point functions of the form 
$\langle e^{iq(\phi(z)+\bar\phi(\bar z))}
e^{iq'(\phi(w)+\bar\phi(\bar w))}\rangle$ in the limit where the bulk
operators approach the boundary, but here $q,q'\in \mathbb{R}$ are 
generic momenta, so these bulk two-point functions can not be computed by 
the method of rotated images. Similarly, the boundary OPE coefficients 
$C_{ijk}$ for general momenta will in general depend on $g$ in a way that
cannot be determined by our $SU(2)$ methods.

\subsection{Mixed bulk-boundary correlation functions}
\label{sec:mixedcorrel}

It is also of interest to compute mixed correlation functions involving
both bulk and boundary fields.  At generic momentum the best we can do is
to once again use conformal invariance to constrain low order amplitudes.  
For a two-point function involving a 
boundary field at $x$ and a bulk field at $z$ the dependence on $x$,
$z$ and $\bar z$ is completely determined by the boundary and bulk scaling
dimensions of the respective fields. We have derived how the 
boundary scaling dimensions depend on the coupling $g$ but the two-point
function also contains the relevant bulk-to-boundary OPE coefficient, 
which involves $g$ in a way that we have not determined.

Mixed correlation functions that involve only discrete fields, both in the 
bulk and on the boundary, can be computed by treating the discrete boundary 
fields as limits of purely holomorphic bulk discrete fields, as described 
in section~\ref{sec:intboundaryfields}, 
\begin{align}
\Psi_{JM}(x) = \lim_{w\rightarrow x}\psi_{JM}(w)\psi_{00}(\bar w) \,,
\label{firstdef}
\end{align}
and using the method of rotated images. 
Alternatively, we can define the boundary field $\Psi_{JM}(x)$ as the
limit obtained when a purely anti-holomorphic bulk discrete field approaches 
the boundary,
\begin{align}
\hat\Psi_{JM}(x) = \lim_{\bar w\rightarrow x}
\psi_{00}(w)\psi_{JM}(\bar w) \,.
\label{seconddef}
\end{align}
In the free theory these two definitions lead to identical results but
this is no longer the case in the interacting theory. This is illustrated
by two-point functions involving one boundary field and one in the bulk. 
Consider for example 
$\langle \Psi_{10,11}(z,\bar z) \Psi_{11}(x)\rangle$, which vanishes by
momentum conservation in the free theory. If we use \eqref{firstdef} and
implement the method of rotated images by moving the integration contours
of the boundary currents into the lower half-plane we find
\begin{align}
\label{onehand}
\langle \Psi_{10,11}(z,\bar z) \Psi_{11}(x)\rangle_g
&= \lim_{w\rightarrow x} \sum_{\bar m=-1}^1 
\mathcal{D}^{1}_{1,\bar m}(g) \,
\langle \psi_{10}(z)\psi_{1\bar m}(z^*) \psi_{11}(w)\rangle \nonumber\\
&= \frac{\sqrt{2}\sin^2\pi g}{(z-z^*)\vert z-x\vert^2} \,.
\end{align}
It is easily checked that the end result is the same when the integration
contours are moved into the upper half-plane. If we instead choose 
\eqref{seconddef} as our boundary field the two-point function vanishes,
as is easily seen by moving the boundary integration contours into the
upper half-plane,
\begin{align}
\label{otherhand}
\langle \Psi_{10,11}(z,\bar z) \hat\Psi_{11}(x)\rangle_g
&= \lim_{w^*\rightarrow x} \sum_{m=-1}^1 
\mathcal{D}^{1}_{0,m}(-g) \,
\langle \psi_{1m}(z)\psi_{11}(z^*) \psi_{11}(w^*)\rangle \nonumber\\
&= 0 \,.
\end{align}
Each term in the sum is zero by momentum conservation in the free 
chiral theory. 

The difference between \eqref{onehand} and \eqref{otherhand} is not
a sign of any pathology in the theory but rather serves as a reminder
that the two definitions \eqref{firstdef} and \eqref{seconddef} select
different bases for the discrete boundary operators, which are related
to one another by a global $SU(2)$ rotation. Boundary amplitudes
of discrete operators are independent of the choice of basis but 
the bulk-to-boundary OPE coefficients are not and this is reflected
in mixed amplitudes involving bulk and boundary discrete fields.

\section{Discussion}
\label{conclusions}
\psfrag{pp}{$e^{ip(z)}$}
\psfrag{mp}{$e^{-ip(w)}$}
\psfrag{ppb}{$e^{ip(z^*)}$}
\psfrag{mpb}{$e^{-ip(w^*)}$}
\EPSFIGURE[h!]{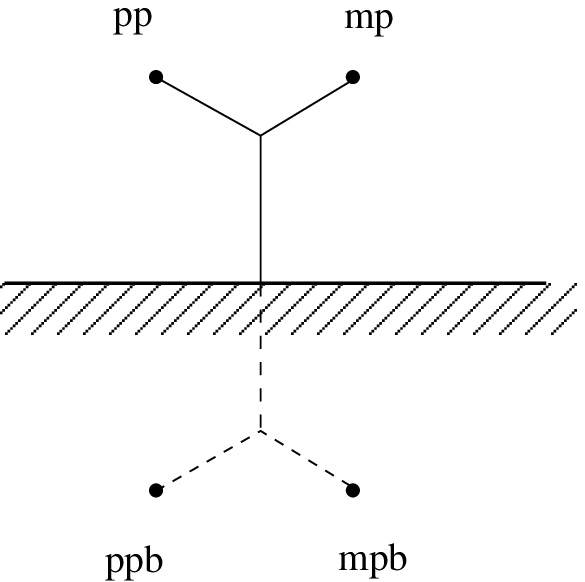, width=5cm}{
Method of rotated images applied in the crossed channel for bulk operators 
with opposite momenta.
\label{fig:oppositemom}}
In this paper we set out to compute correlation functions of primary 
fields in the two-dimensional boundary conformal field theory of a scalar 
field interacting with a critical periodic boundary potential.  
By extending the method of rotated images introduced in \cite{Callan:1994ub}
we have shown how to calculate general correlation functions of both bulk 
and boundary primary fields when the boson is compactified at the self-dual
radius. At other allowed boson radii our methods  
still generate all correlation functions of the discrete fields, which carry
integer multiples of the lattice momentum defined by the periodic interaction,
but not for fields with fractional momenta compared to the lattice momentum.
In particular, we would like to compute correlation functions for primary
fields carrying continuum values of the momentum in the theory at infinte
boson radius.  

It should in principle be possible to use the method of rotated images to
calculate bulk two-point functions of operators with generic momenta.  
These are expected to be non-vanishing whenever the two momenta add up to
a lattice momentum, but let us focus on the simplest case of opposite
momenta.  One would start by reflecting the anti-holomorphic fields
in the two-point function 
\begin{equation}
\langle e^{ip\phi(z)}e^{ip\bar\phi(\bar z)}
e^{-ip\phi(w)}e^{-ip\bar\phi(\bar w)}\rangle \,,
\label{generictwopoint}
\end{equation}
through the real axis and then apply the OPE of the free chiral theory to 
the holomorphic image fields in the lower half-plane. 
The fields on the right-hand side of this OPE all carry 
momentum zero and can therefore be written as linear combinations of 
discrete primary fields with $m=0$ and descendants of these primaries.  
The method of rotated images is then applied term by term, rotating 
descendants in the same representation as the corresponding primaries.  
One then also writes out the OPE of the two holomorphic fields in 
\eqref{generictwopoint} and finally carries out the sum over the 
remaining chiral two point functions. 
What makes this approach challenging is the non-trivial decomposition of 
operators of the type $\partial^{n_1}\phi\cdots\partial^{n_k}\phi$ into 
different $J$-families.  We have calculated the first few terms but in 
order to make use of this method one really needs to find the exact sum 
of the series, or in other words work out the exact conformal block
that encodes the $SU(2)$ rotation on the image fields.
It would be interesting to see how the $g$-dependence of the scaling
dimension of boundary operators arises in such an approach.

\acknowledgments

During the course of this work we have benefitted from discussions with 
C.\ Callan, S.\ Chaudhuri, D.\ Friedan, J.\ McGreevy, and I.\ Klebanov.
This work was supported in part by 
the Institute for Theoretical Physics at Stanford University, 
by NORDITA the Nordic Institute for Theoretical Physics, 
and by grants from the Icelandic Research Council, 
The University of Iceland Research Fund, and The Icelandic 
Research Fund for Graduate Students.

\appendix

\section{Rotation coefficients}
\label{app:rotation}

In this appendix we quote a general formula \cite{Hamermesh:1962} for the 
rotation coefficients $\mathcal{D}^j_{m,n}(g)$ in \eqref{eq:coeff} and 
give explicit examples for low values of $j$. For simplicity we 
take $g = \bar g$. In this case the rotation coefficients can 
all be expressed in terms of $\sin\pi g$ and $\cos\pi g$ as follows
\begin{align}
\mathcal{D}^j_{m,n}(g) = \sum_\ell
\frac{\sqrt{(j{+}m)!(j{-}m)!(j{+}n)!(j{-}n)!}}{
(j{-}m{-}\ell)!(j{+}n{-}\ell)!\ell!(m{-}n{+}\ell)!}
(\cos\pi g)^{2j-2\ell+n-m}(i\sin\pi g)^{2\ell+m-n}.
\end{align}
where the sum is over integers $\ell$ and the range of summation
is cut off when the argument in any one of the factorials in 
the denominator goes negative. When $g = \bar g$ the rotation 
coefficients are symmetric, $\mathcal{D}^j_{m,n}(g) 
= \mathcal{D}^j_{n,m}(g)$.

The representation for $j=0$ is trivial, $\mathcal{D}^0_{0,0} = 1$.
In the $j=1/2$ representation we have
\begin{equation}
\mathcal{D}^{1/2}(g) = 
\left(
\begin{array}{cc}
\cos\pi g & i \sin\pi g \\
i \sin\pi g & \cos\pi g\\
\end{array}\right)\,,
\end{equation}
and for $j=1$ we find
\begin{equation}
\mathcal{D}^{1}(g) = 
\left(
\begin{array}{ccc}
\cos^2\pi g & \frac{i}{\sqrt{2}} \sin2\pi g & -\sin^2\pi g\\
\frac{i}{\sqrt{2}} \sin2\pi g & \cos2\pi g & \frac{i}{\sqrt{2}} \sin2\pi g
\\
-\sin^2\pi g & \frac{i}{\sqrt{2}} \sin2\pi g & \cos^2\pi g\\
\end{array}\right) \,.
\end{equation}
The extension to higher values of $j$ is straightforward.

The rotation coefficients satisfy a number of useful relations
including 
\begin{align}
\mathcal{D}^j_{m,n}(-g)&= (-1)^{n-m}\,\mathcal{D}^j_{-n,-m}(g) \,, \\
\mathcal{D}^{j_1}_{m_1,n_1}(g)\mathcal{D}^{j_2}_{m_2,n_2}(g)
&= \sum_{J=\vert j_1-j_2 \vert}^{j_1+j_2}
C^{J (m_1+m_2)}_{j_1 m_1,j_2 m_2}
C^{J (n_1+n_2)}_{j_1 n_1,j_2 n_2}
\mathcal{D}^{J}_{(m_1+m_2),(n_1+n_2)}(g) \,,
\end{align}
where $C^{JM}_{j_1 m_1,j_2 m_2}$ are Clebsch-Gordan 
coefficients.

\section{Open string spectrum from closed string channel}
\label{app:closedchannel}
In this appendix we give an alternative derivation of the spectrum
\eqref{eq:generalspectrum} of open strings with endpoints interacting 
with two different boundary potentials.
The calculation is a straightforward generalization of that presented
in the ``Note added'' at the end of \cite{Callan:1994ub} and, for the
most part, we use their notation. 

We consider a non-compact boson on a strip of width $\ell$ with different 
boundary potential strengths $g_1,g_2\in\mathbb{R}$ at each boundary.
The partition function in the closed string channel is
\begin{equation}
Z = \langle B,g_1 \vert q^{L_0+\tilde L_0} \vert B,g_2\rangle \,,
\end{equation}
where $q=e^{-2\pi\ell/\beta}$ with $\beta$ the parameter length of the closed 
string, and $\vert B,g\rangle$ is the boundary state induced by the 
boundary interaction \eqref{bpotential}.  The exact expression for
this boundary state, obtained in \cite{Callan:1994ub}, is given 
by\footnote{Our normalization convention for $\vert B,g\rangle$ differs 
from the one adopted in \cite{Callan:1994ub}.}
\begin{equation}
\vert B,g\rangle = \frac{1}{\sqrt{2\pi}} \sum_{j=0,\half,1\ldots}
\sum_{m=-j}^j \mathcal{D}^j_{m,-m}(g)\,\vert j,m,m \rangle\rangle \,,
\end{equation}
where $\vert j,m,m' \rangle\rangle$ denotes the reparametrization
invariant Ishibashi state \cite{Ishibashi:1988kg} based on the bulk 
discrete primary field $\Psi_{j,m;j,m'}(z,\bar z)$. 

The partition function is thus given by
\begin{equation}
Z = \frac{1}{2\pi} \sum_{j_1,m_1} \sum_{j_2,m_2} 
\mathcal{D}^{j_1}_{m_1,-m_1}(g_1)^*\,
\mathcal{D}^{j_2}_{m_2,-m_2}(g_2)
\langle\langle j_1,m_1,m_1\vert q^{L_0+\tilde L_0} \vert
j_2,m_2,m_2\rangle\rangle .
\end{equation}
The sum over descendants in the Ishibashi states produces a Virasoro
character,
\begin{equation}
Z = \frac{1}{2\pi} \sum_{j=0,\half,1,\ldots} \chi_j^\textrm{Vir}(q^2)
\sum_{m=-j}^j \mathcal{D}^{j}_{m,-m}(g_1)^*\,
\mathcal{D}^{j}_{m,-m}(g_2) \,,
\label{bfour}
\end{equation}
leaving us with a sum over rotation coefficients that can
be converted to an integral over $SU(2)$ characters as follows,
\begin{align}
\sum_{m=-j}^j \mathcal{D}^{j}_{m,-m}(g_1)^* \,
\mathcal{D}^{j}_{m,-m}(g_2) 
=& \sum_{m=-j}^j \langle j,m\vert e^{2\pi ig_2J_1}\vert j,-m\rangle
\langle j,-m\vert e^{-2\pi ig_1J_1}\vert j,m\rangle \nonumber \\
=& \int_{-\pi}^\pi \frac{d\phi}{2\pi} \sum_{m=-j}^j
\langle j,m\vert e^{i\phi J_3} e^{2\pi ig_2J_1} e^{-i\phi J_3}
e^{-2\pi ig_1J_1}\vert j,m\rangle \nonumber \\
=& \int_{-\pi}^\pi \frac{d\phi}{2\pi}\,
\frac{\sin (2j+1) \alpha/2}{\sin \alpha/2} \,, \nonumber \\
=& \int_{-\pi}^\pi \frac{d\phi}{2\pi}\,
\chi_j^{SU(2)}(\alpha) \,.
\label{bfive}
\end{align}
where the net rotation angle $\alpha$ can, for example, be determined 
using the $j=1/2$ representation of $SU(2)$ for which $J_i=\half\sigma_i$, 
with $\sigma_i$ Pauli matrices. After some straightforward algebra one
finds
\begin{equation}
\sin^2(\alpha/4) = \sin^2(\pi g_-) \cos^2(\phi/2)
+ \cos^2(\pi g_+) \sin^2(\phi/2) \,,
\end{equation}
where $g_\pm=\half(g_1\pm g_2)$. This reduces to \eqref{eq:generalspectrum} 
if we make the identifications $\alpha=4\pi\lambda$ and $\phi=2\pi k$.
Finally, we can rewrite the partition function $Z$ in the form 
\eqref{partfcn} by inserting \eqref{bfive} into \eqref{bfour} and then 
converting to the open string channel by the following modular 
transformation 
\begin{equation}
\sum_{j=0,\half,1,\ldots}  \frac{1}{\sqrt{2}} \,
\chi_j^{SU(2)}(\alpha) \,
\chi_j^\textrm{Vir}(q^2) 
= \frac{1}{\eta(\omega)} \sum_{m=-\infty}^\infty 
\omega^{(m+\frac{\alpha}{4\pi})^2} \,.
\end{equation}

\bibliographystyle{JHEP}
\bibliography{refs}        

\end{document}